\documentclass[useAMS,usenatbib,usegraphicx]{mn2e}

\title[Aligned grains in massive YSOs]{Aligned grains and inferred toroidal magnetic fields in the envelopes of massive young stellar objects\thanks{Based on observations made with the NASA/ESA Hubble Space Telescope, 
obtained at the Space Telescope Science Institute, which is operated by the Association 
of Universities for Research in Astronomy, Inc., under NASA contract NAS 5-26555.
These observations are associated with programme \#10519.}
}

\author[J. P. Simpson et al.]
{Janet P. Simpson$^{1}$\thanks{E-mail: jsimpson@seti.org},
Barbara A. Whitney$^{2}$, Dean C. Hines$^{3}$, Glenn Schneider$^4$, 
\newauthor
Michael G. Burton$^{5}$, Sean W. J. Colgan$^{6}$, Angela S. Cotera$^{1}$, Edwin F. Erickson$^{6}$, 
\newauthor
Michael J. Wolff$^{7}$  \\
$^{1}$SETI Institute, 189 Bernardo Ave, Mountain View, CA 94043, USA\\
$^{2}$University of Wisconsin, 475 N. Charter St., Madison, WI 53706, USA\\
$^{3}$Space Telescope Science Institute, 3700 San Martin Dr., Baltimore, MD 21218, USA\\
$^{4}$University of Arizona, Steward Observatory, 933 N. Cherry Ave., Tucson, AZ 85721, USA\\
$^{5}$School of Physics, University of New South Wales, Sydney, NSW 2052, Australia\\
$^{6}$NASA Ames Research Center, Moffett Field, CA 94035, USA \\
$^{7}$Space Science Institute, 4750 Walnut St., Suite 205, Boulder, CO 80301, USA }

\begin{document}
\date{}
\pagerange{\pageref{firstpage}--\pageref{lastpage}} \pubyear{2013}

\maketitle

\label{firstpage}

\begin{abstract}

Massive young stellar objects (YSOs), like low-mass YSOs, 
are thought to be surrounded by optically thick envelopes and/or discs and
are observed to have associated regions that produce polarized light 
at near-infrared wavelengths.
These polarized regions are thought to be lower-density outflows along the polar axes 
of the YSO envelopes. 
Using the 0.2-arcsec spatial resolution of the Near-Infrared Camera and Multi-Object Spectrometer
on the {\it Hubble Space Telescope} 
we are examining the structure of the envelopes and outflow regions of massive YSOs 
in star-forming regions within a few kpc of the Sun. 
Here we report on 2-\micron\ polarimetry of Mon R2-IRS3, S140-IRS1, and AFGL 2591.

All three sources contain YSOs with highly-polarized monopolar outflows, 
with Mon R2-IRS3 containing at least two YSOs in a small cluster.  
The central stars of all four YSOs are also polarized, 
with position angles perpendicular to the directions of the outflows. 
We infer that this polarization is due to scattering and absorption by aligned grains.
We have modelled our observations of S140-IRS1 and AFGL 2591 
as light scattered and absorbed both by 
spherical grains and by elongated grains that are aligned by magnetic fields. 
Models that best reproduce the observations have 
a substantial toroidal component to the magnetic field 
in the equatorial plane. 
Moreover, the toroidal magnetic field in the model that best fits AFGL 2591 
extends a large fraction of the height of the model cavity, which is $10^5$ au.
We conclude that the massive YSOs in this study 
all show evidence of the presence of a substantial toroidal magnetic field.  

\end{abstract}

\begin{keywords}
stars: massive --
stars: protostars --
ISM: magnetic fields --
ISM: jets and outflows --
infrared: ISM --
infrared: stars 
\end{keywords}

\section{Introduction}

Although there has been substantial progress in understanding massive star formation 
in recent years (see, e.g., Bonnell \& Smith 2011 for a review), 
there remain many interesting questions.
One such concern is the influence of magnetic fields, 
which are known to exist in molecular clouds where stars form.
Shu, Adams, \& Lizano (1987) first described how gravity could collapse a low-mass cloud 
along magnetic field lines to form a disc and then a star via ambipolar diffusion. 
Here a disc is defined as `a long-lived, flat, rotating structure in centrifugal equilibrium,' 
(Cesaroni et al. 2007).
However, for massive stars, 
Allen, Li, \& Shu (2003), Hennebelle \& Fromang (2008), and Mellon \& Li (2008)
have shown that magnetic braking due to even a weak magnetic field
inhibits the formation of a thin, rotationally supported `disc', 
leaving only a non-Keplerian `pseudodisc' (Galli \& Shu 1993).
On the other hand, Seifried et al. (2012) find 
that when turbulence is included in their simulations, 
Keplerian discs are formed in spite of magnetic braking. 

Irrespective of the presence of a true disc, 
the effect on the magnetic field lines is that the poloidal component 
is pinched towards the centre in the equatorial plane.  
In addition, a toroidal component can appear in the disc, pseudodisc, or 
disc-like toroid, 
here described as a condensation in the equatorial plane that is not quite yet a `disc'
(e.g., Hennebelle \& Fromang 2008; Seifried et al. 2011).
Such pinches in the magnetic field lines have been found in the envelopes
of both low and high mass young stellar objects 
(YSOs), for example, by Girart et al. (2006, 2009), who measured the 
sub-mm wavelength polarized emission from aligned grains in such objects.

On the other hand, evidence for a toroidal field in YSO discs or toroids is not so convincing.
Since grain alignment by magnetic fields is thought to be generally true 
(see Lazarian 2007, 2009, for reviews of grain alignment physics), 
the presence of a toroidal field should also be discernible through 
polarization measurements of 
the position angle (PA) of the magnetic field in the plane of the sky.

For most stars and YSOs, the magnetic field causing grain alignment is 
the Galactic interstellar magnetic field. 
It has long been known that visible-wavelength stellar polarization 
is correlated with the features seen in the Galactic plane 
(see e.g., Heiles \& Crutcher 2005 and references therein); 
currently Clemens et al. (2012) are extending these studies to much farther distances 
in the Galactic plane through near-infrared (NIR) observations of stellar polarization.

For visible to mid-infrared (MIR, $\sim 8 - 25$ \micron) wavelengths, the polarization from
non-spherical, aligned grains is a complicated function of scattering, absorption, 
and magnetic field direction (Whitney \& Wolff 2002; Smith et al. 2000).
Scattering by aligned grains produces polarization whose PA 
is a combination of the vectors perpendicular to the magnetic field direction 
and perpendicular to 
the `scattering plane', defined by the photon source, the scattering particle, and the observer. 
On the other hand, absorption by aligned grains produces polarization vectors 
parallel to the magnetic field direction.
Either effect (scattering or absorption) can dominate, depending on field direction 
and optical depths (see Whitney \& Wolff 2002 for examples).

For MIR to sub-mm wavelengths, polarization perpendicular to the magnetic field 
in the plane of the sky is produced by emission from aligned grains
(scattering is negligible).
This polarization tracks the magnetic field in molecular clouds and YSO envelopes 
(e.g., Dotson et al. 2000;  Girart et al. 2006, 2009; Curran \& Chrysostomou 2007). 
Aitken et al. (1993) and Wright (2007) have plotted the orientation of YSO polarization 
versus the local Galactic magnetic field direction and YSO outflow direction, 
mostly using their MIR spectra from the atlas of Smith et al. (2000).
There is no special correlation of source polarization PA with outflow direction. 
A significant problem could be that their aperture on the source includes 
both the YSO disc and the outflow, which may have perpendicular polarization 
orientations (Aitken et al. 1993; Wright 2007). 

Magnetic field directions, however, cannot be estimated from the polarization of spherical grains. 
In this case, scattering tends to produce polarization vectors aligned perpendicular 
to the scattering plane. 
As a result, an optically thin nebula of non-aligned grains 
illuminated by a central source shows a circular pattern of polarization vectors.

Since it is essential to understand 
the relation between massive YSOs' outflows and discs,
we have undertaken a study of such systems with the {\it Hubble Space Telescope} ({\it HST}), 
using the NIR polarimetry capability of 
its Near-Infrared Camera and Multi-Object Spectrometer (NICMOS).
For this study we chose the YSOs and candidate YSOs closest to the Earth 
whose luminosities indicate that their masses are $> 8$ M$_\odot$
(log L/L$_\odot \ga 3.5$). 
We have previously described three objects that are seen essentially edge-on:
NGC 6334-V, and NIRS1 and NIRS3 in S255-IRS1 (Simpson et al. 2009).
In this paper we report on three more sources containing massive YSOs: 
Mon R2-IRS3, S140-IRS1, and AFGL 2591. 
Our goals are to  
characterize the structure of any circumstellar discs and outflow regions  
and to determine the orientation of the local magnetic field 
in the plane of the sky.

Mon R2 is a cluster of \mbox{H\,{\sc ii}} regions and YSOs, 
where IRS2 is the illuminating star of the shell-like \mbox{H\,{\sc ii}} region IRS1 
(e.g., Howard, Pipher, \& Forrest 1994; Aspin \& Walther 1990; Yao et al. 1997)
and IRS3 is a luminous cluster of YSOs and other stars 
(Beckwith et al. 1976; Preibisch et al. 2002; Alvarez et al. 2004a).
Although the components of IRS3 are not resolved at 24.5 \micron\ 
when observed with a resolution of 0.6 arcsec (de Wit et al. 2009), 
recent interferometry at 10 \micron\ indicates additional structures 
that may be due to the presence of a circumstellar disc (Linz et al. 2011). 
The distance is estimated to be 830 pc (Herbst \& Racine 1976).
The total luminosity of the compact group is $\sim 1.4 \times 10^4$ L$_\odot$ 
(Henning, Chini, \& Pfau 1992). 

S140-IRS1 is the most luminous of a number of infrared sources 
located in the L1204 molecular cloud that provides a bright rim to the S140 \mbox{H\,{\sc ii}} region. 
We assume that the distance is that of the L1204 cloud source IRAS 22198+6336,
which has a parallax distance of $\sim 764$ pc (Hirota et al. 2008).
The morphology of the red-shifted and blue-shifted lines in the outflow 
are consistent with the pole of the outflow being close to the line of sight 
(Minchin, White, \& Padman 1993). 
The centrosymmetric NIR polarization vectors also indicate that the outflow cavity 
is close to the line of sight (Joyce \& Simon 1986).
However, the positioning of the outflow components in the plane of the sky 
is sufficiently uncertain that Trinidad et al. (2007), Preibisch \& Smith (2002), 
Weigelt et al. (2002), and Yao et al. (1998) suggested that there are  
multiple outflows in different directions perhaps caused by multiple YSOs.
This may be possible, as the appearance at NIR, MIR (de Wit et al. 2009), 
and radio wavelengths is very clumpy. 

AFGL 2591, at a distance of 3.33 kpc (Rygl et al. 2012), is the most massive and 
most luminous YSO in our current data set.
Until Rygl et al.'s (2012) recent measurement, it was thought that the distance was much 
smaller, $\sim 1$~kpc; consequently many of the source parameters in the literature need revisiting.
This source has an extended CO outflow (Lada et al, 1984; Hasegawa \& Mitchell 1995), 
H$_2$ outflows (Poetzel, Mundt, \& Ray 1992; Tamura \& Yamashita 1992), and 
numerous H$_2$O (Trinidad et al. 2003; Sanna et al. 2012)
and OH masers (Hutawarakorn \& Cohen 2005).
In fact, the numbers of masers and radio continuum sources in the vicinity indicate 
that AFGL 2591 is only the most massive of a cluster of young stars and YSOs 
(Trinidad et al. 2003; Sanna et al. 2012).
A three-colour (JHKs) Gemini image by C. Aspin is published as fig. 16 in Zinnecker \& Yorke (2007);
this shows us that only the blue-shifted lobe of the outflow is visible at NIR wavelengths 
and that this outflow has intriguing loops or rings to the west of the YSO 
(e.g., Minchin et al. 1991; Preibisch et al. 2003).

All three sources include other stars in our fields of view. 
This almost certainly indicates that they are parts of clusters, 
as is predicted by theories of massive star formation (e.g., Zinnecker \& Yorke 2007).

In this paper we describe our observations in Section 2 and the results in Section 3. 
We include polarimetry of the other stars in the field of view 
as well as the YSOs and their scattered light outflow regions.
In Section 4 we present radiation transfer models and discuss how they compare to 
the observations. 
In Section 5 we compare our observations to other observations 
and describe some numerical simulations of massive star formation in the literature 
that include the presence of magnetic fields. 
Finally, in Section 6 we present our summary and conclusions. 

\section{Observations}

\subsection{NICMOS Polarization Data}


\setcounter{table}{0}
\begin{table*}
\centering
\begin{minipage}{80mm}
\caption{Journal of the observations.}
\begin{tabular}{@{}cccccc@{}}
\hline
Visit & Date & Target & Center RA & Center Dec & Camera 2 \\ 
      &      &        & (J2000)  &  (J2000) & Position Angle \\ 
      &      &        &          &          & (degrees) \\
\hline
1  & 2006 Aug 31 & Mon R2-IRS3 & 06:07:47.84 & $-$06:22:56.29 & 32.66   \\
54  & 2006 Oct 8 & Mon R2-IRS3 & 06:07:47.84 & $-$06:22:56.29 & 59.57   \\
7 & 2006 May 17 & AFGL 2591 & 20:29:24.89 & +40:11:19.60 & 14.57   \\ 
8 & 2006 Jul 1 & AFGL 2591 & 20:29:24.89 & +40:11:19.60 & $-$25.43   \\ 
9 & 2006 Apr 22 & S140-IRS1 & 22:19:18.32 & +63:18:45.40 & 67.57  \\
10 & 2006 Mar 30 & S140-IRS1 & 22:19:18.32 & +63:18:45.40 & 96.44  \\
11 & 2006 Apr 10 & Oph-N9 & 16:27:13.31 & $-$24:41:32.40 & 68.88 \\

\hline
\end{tabular}
\end{minipage}
\end{table*}


\begin{figure}
\includegraphics[width=84mm]{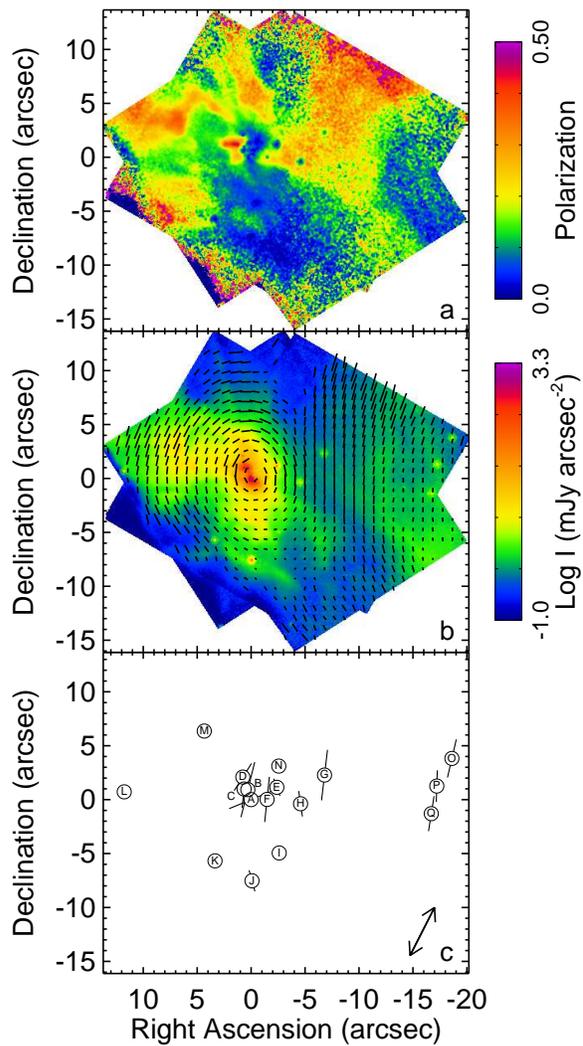}
\caption{
{\it HST} NICMOS image of Mon R2-IRS3. 
Panel (a): fractional polarization.
Panel (b): log intensity with polarization vectors.
Panel (c): locations of the stars detected with NICMOS (Table 2). 
The lines drawn through the stars with statistically significant polarization 
are proportional to $P$ plus a constant and plotted at the measured 
polarization PA $\theta$. 
Note that the west side of the image, including the three stars, 
is located in IRS2. 
The double arrow marks the direction of the Galactic plane.
}
\end{figure}

\begin{figure}
\includegraphics[width=84mm]{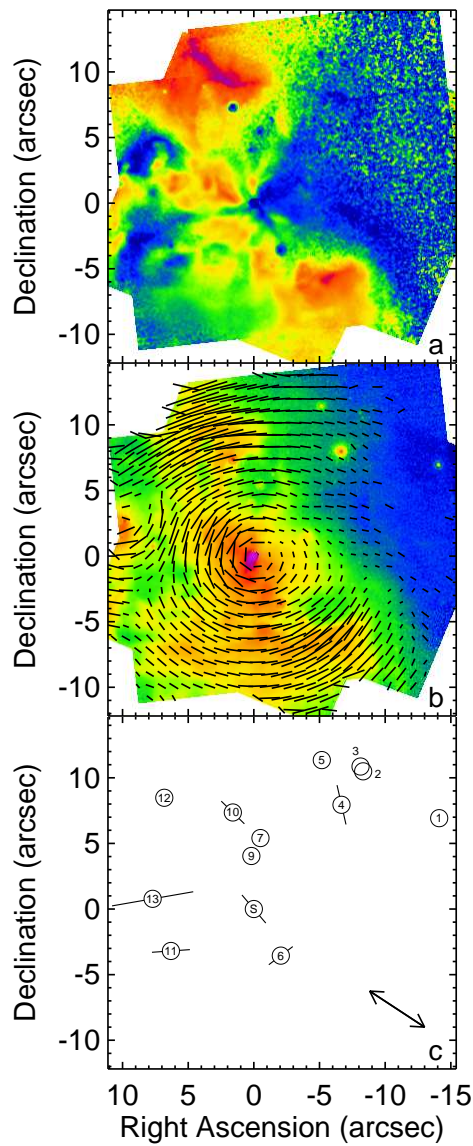}
\caption{
{\it HST} NICMOS image of S140-IRS1. 
The source off the edge of the image at approximately +10,+3 arcsec is IRS3.
Panel (a): fractional polarization.
Panel (b): log intensity with polarization vectors.
Panel (c): locations of the stars detected with NICMOS (Table 3). 
The position of IRS1 is marked with the letter `S'.
See Fig. 1 for a description of the lines in this panel.
}
\end{figure}

\begin{figure}
\includegraphics[width=84mm]{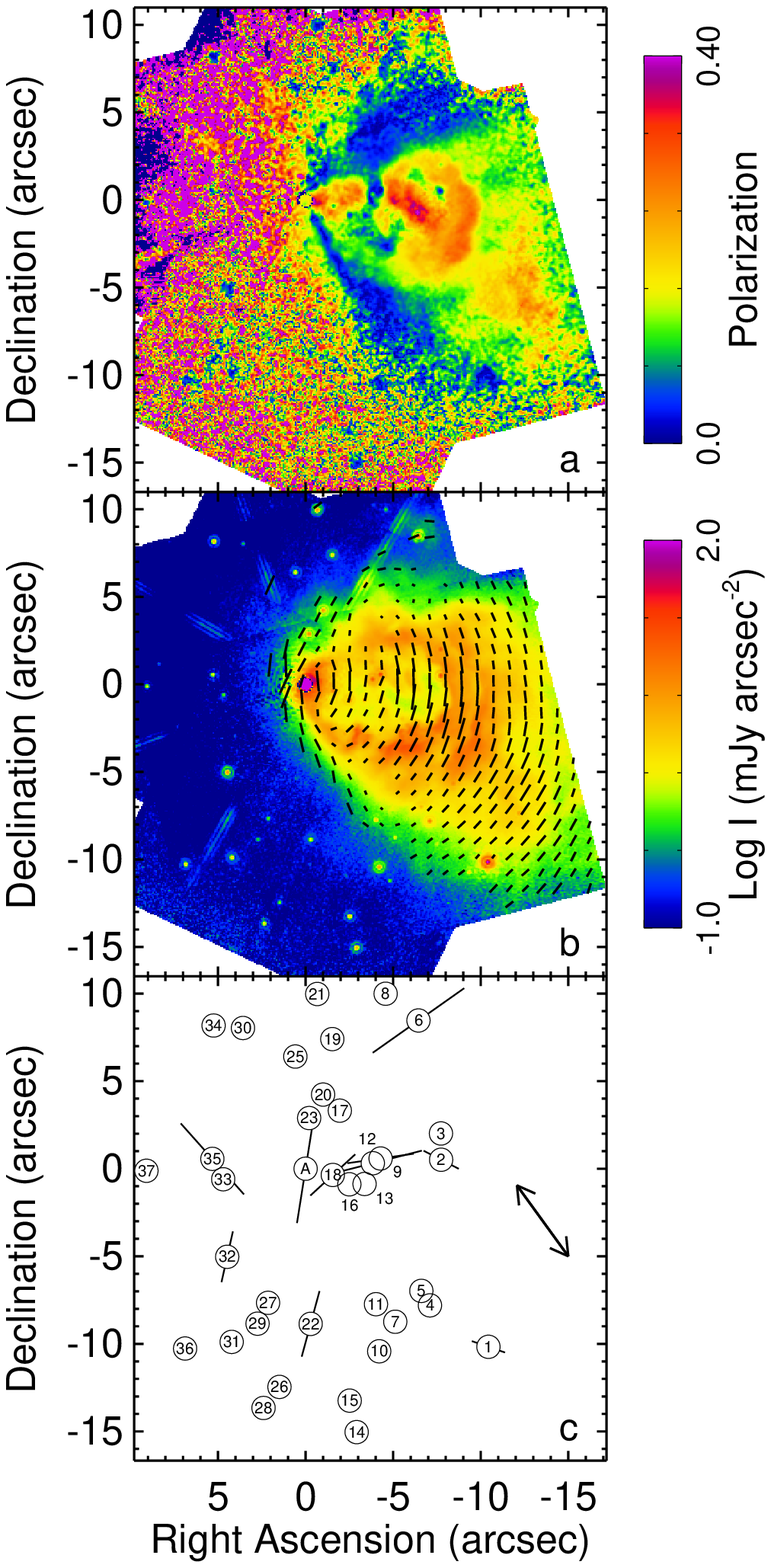}
\caption{
{\it HST} NICMOS image of AFGL 2591. 
Panel (a): fractional polarization.
Panel (b): log intensity with polarization vectors.
The {\it HST} diffraction pattern is visible in those parts of the mosaicked image 
that were observed with the YSO centred in the array and not positioned behind 
the coronagraph hole.
Panel (c): locations of the stars detected with NICMOS (Table 4). 
The position of AFGL 2591 is marked with the letter `A'.
See Fig. 1 for a description of the lines in this panel.
}
\end{figure}

We observed each source for two visits with NICMOS on {\it HST} 
with the Camera 2 POL0L, POL120L, and POL240L filters 
(hereafter the `POL filters').
These filters cover a 1.9 -- 2.1 \micron\ bandpass with 0.2-arcsec resolution. 
Table~1 contains a journal of the observations 
(there was originally a visit 4 but it failed and was replaced with visit 54).
In addition, we observed the red standard star Oph-N9
($Ks = 9.620$, $H - K = 2.862$, Persson et al. 1998).
Oph-N9, also known as GY232 (Greene \& Young 1992) and
BKLT 162713$-$244133 (Barsony et al. 1997),
is a highly extincted giant behind the $\rho$ Oph cloud (Luhman \& Rieke 1999).

The purpose for having two visits per source was so that the 
orientations of the Camera 2 array with respect to north could be 
substantially different, 
with the result that the effects of the systematic errors 
caused by the non-optimum NICMOS polarization filters are reduced when the visits are averaged.
Unfortunately, in Cycle 14 {\it HST} had to be operated in `two-gyro' mode,
which limited the visit length for most orientations as a function of location on the sky
(in particular, visit 8 was shorter than the other visits). 
Larger differences in orientation than were obtained (Table~1) are preferable,
but they were not possible without reducing the visit length to very short, unusable times.

Because the YSOs observed in this paper are so bright, 
their point spread functions (PSFs) extend over most of the field of view.
This is a particular problem when the YSO is polarized because its polarized PSF 
affects the measured polarization of the nebulosity of interest.
Therefore, to improve PSF subtraction in each visit we measured each source 
in two separate sequences, 
once with a four-position spiral dither pattern with spacing 1.0213~arcsec 
and once with the YSO centred in the NICMOS Camera 2 coronagraph hole.
For the spiral dither pattern, the detector array was read out 
in MULTIACCUM mode
with sample sequence STEP8 to accumulate total times 
ranging from 32 to 56 s per dither position per filter.
In the coronagraph mode the detector array was read out, also in MULTIACCUM mode,
with sample sequence STEP2 for the YSOs and STEP8 for Oph-N9 
to achieve total integration times of 207 s per POL filter
for the YSOs and 280 s for Oph-N9.
 
The data were reduced using the procedure described by Simpson et al. (2009).
This includes dark subtraction, flat fielding, correction for the electronic ghosts 
due to amplifier ringing 
(also known as the `Mr. Staypuft' anomaly, Thatte et al. 2009), 
bad pixel correction, and correction for the `pedestal' effect. 
For the coronagraph images, specialized bad pixel masks were used to 
compensate for the lack of dithering (G. Schneider, in preparation).
None of the images showed the noticeable jumps from quadrant to quadrant 
that are usually ascribed to the pedestal effect;  
however, when the minimum flux was computed for each image by measuring the median
of $\sim 800$ pixels centred on exactly the same position on the sky, 
it was found that these minima are never the same. 
Consequently, small constants ($\sim 0.01 - 0.11$ counts s$^{-1}$)
were subtracted from the dithered images with the larger minimum fluxes
so that all images would have the same flux levels for later median combining.
These variations could be due to pedestal effect contributions. 
Estimates of the {\it HST} thermal background were also subtracted. 
The result of this analysis is an uncertainty $\la 0.03$ counts s$^{-1}$
in the continuum level for each image,  
but this corresponds to only 0.016 mJy arcsec$^{-2}$, 
substantially fainter than any of the nebulosity that we are measuring
(typically the minimum of the intensities that have adequate signal-to-noise ratios for 
polarization measurement is $\sim 0.3$ mJy arcsec$^{-2}$). 

For each polarizing filter, we aligned and shifted the dither positions
for the YSO-centred images 
by centroiding two bright stars using the Interactive Data Language ({\sc idl}) program, 
{\sc idp3} (Stobie \& Ferro 2006).
The shifted images (with the very small additions/subtractions to the flux described above)
were then median-combined to remove the remaining bad pixels;
the result is three images for the three POL filters in each visit, 
all aligned to the same position on the sky. 
The coronagraph images were treated similarly -- median-combined and shifted 
to the same registration as the YSO-centred images by aligning the 
other stars in the field.

Even with the use of the coronagraph, there is a significant PSF due to the 
occulted YSO that must be removed before the images in a single POL filter can be combined.
Since the NICMOS POL filters have a 10 percent bandpass, 
the colour of the star has an important effect on the measured PSF.
Because our YSOs are very red ($H - K$ ranging from 2.9 to 4.2),
it was necessary to find a very red PSF standard that does not have strong 
spectral features at 2 \micron.
Although Oph-N9 is close to being red enough ($H - K = 2.86$), 
it is not very bright, such that its PSF becomes quite noisy greater than 
about 2 arcsec from its centre.
A less red but usable PSF can be found in the bright NIR standard GJ-784 
(HD191849, $K = 4.28$, $H - K = 0.24$)
(Programme 10847, PI: D. Hines; G. Schneider, in preparation). 
Because the red Oph-N9 PSF has distinct features in the region within 10 -- 20 pixels 
from the coronagraph hole 
that we also see in the YSO data, 
we subtracted the Oph-N9 PSF from the coronagraph images 
from 0.5 to 1.5 arcsec and the GJ-784 PSF from all pixels at larger distances 
than 1.5 arcsec.
Since Oph-N9 is polarized (Table~2), the PSFs were subtracted filter by filter
with the normalization constants obtained by measuring the star fluxes in each POL filter.
To enhance the appearance of the figures in this paper, the central 0.6 arcsec 
(radius 4 pixels) 
are replaced by the central image of the YSO;
however, none of the results described hereafter depends on this 
composite image that includes the YSO.
The shifted, PSF-subtracted coronagraph images were then augmented by adding the 
YSO-centred images for those parts of the sky that were not covered by the 
coronagraph images. 
These images were not averaged because the coronagraph images have more integration 
time and because they have the polarized PSF already removed. 

For each visit, the Stokes $I$, $Q$, and $U$ intensities were
computed from the reduced data (Hines et al. 2000; Batcheldor et al. 2006, 2009),
the pixels were rectified to the same plate-scale in both $x$ and $y$ 
(0.075948 arcsec per pixel), and rotated so that north is up. 
The resulting celestially-aligned Stokes
$I$, $Q$, and $U$ images from each pair of visits were then mosaicked
together and the fractional polarization $P$ and the PA $\theta$
of the polarization vectors were computed from the combined $I$, $Q$, and
$U$ using the usual relationships $P = (Q^2+U^2)^{0.5}/I$ and 
$\theta = 0.5\ {\rm arctan} (U/Q)$.
The images used in the mosaics were also smoothed with a $3\times3$ boxcar 
to achieve higher signal/noise for $P$ and $\theta$ images. 
This approximates the {\it HST} spatial resolution of 0.2 arcsec at 2.0 \micron.
Figs. 1 -- 3 show these mosaics, with polarization vectors and fractional polarization. 
The intensity images plotted in this paper are not smoothed
in order to preserve details such as the fine structure of the {\it HST}
diffraction pattern; however, the plotted polarization images and the over-plotted
polarization PA vectors are from the smoothed data.

\subsection{Other NICMOS Data}

\begin{figure}
\includegraphics[width=84mm]{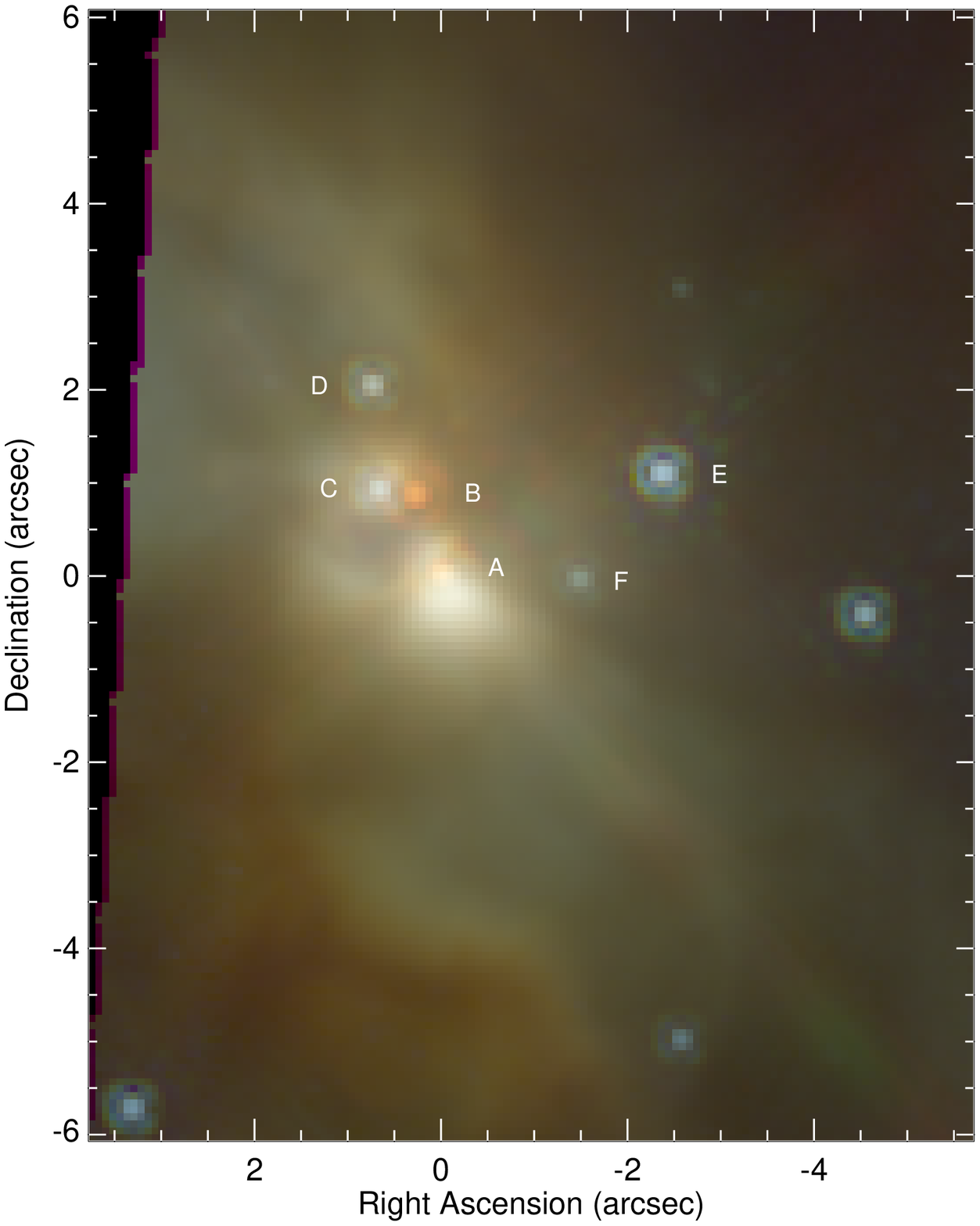}
\caption{
Three-colour {\it HST} NICMOS image of the central cluster of Mon R2-IRS3 
taken with the $F110W$ (blue), $F165M$ (green), and $F207M$ (red) filters. 
The letters identifying the stars are from Preibisch et al. (2002).
}
\end{figure}

In 1997 December the Mon R2 Cluster was imaged with NICMOS on {\it HST}
(Programme 7417, PI: M. Meyer) at 1.10, 1.60, 1.65, and 2.07 \micron.
Andersen et al. (2006) describe their programme and show a colour image of their results.
We downloaded these same Mon R2 data from the {\it HST} archive 
and reduced the frames containing IRS3 and environs
in the same manner as described above for the NICMOS polarization filters.
The results are plotted in Fig. 4 with a log stretch 
(blue is the $F110W$ filter, green is $F165M$, and red is $F207M$).
Unfortunately, these NICMOS images do not include the area of the sky 
at higher right ascension that is seen in the field of view shown in Fig.~1.

Preibisch et al. (2002) also reduced these same NICMOS data from the HST archive; 
their paper includes a table of the coordinates and $J$, $H$, and $K$ magnitudes
of the stars in the IRS3 cluster, 
plus further discussion of the colour-colour diagram.
Fig. 4 includes the star identification letters from Preibisch et al. (2002).

\subsection{Stellar Fluxes and Polarization Measurements}


\setcounter{table}{1}
\begin{table*}
\centering
\begin{minipage}{165mm}
\caption{Stellar polarization and photometry measurements of Mon R2-IRS2 and IRS3 and Oph-N9.
Values for polarization are listed for stars that are detected on the $Q$ and/or $U$ images.
Because of the possibility of systematic uncertainties, the minimum uncertainty in the polarization is 
increased to 1 percent, even if the statistical uncertainty is much smaller.
The coordinates in Mon R2 are derived from the position of Star A in IRS3, as given by Preibisch et al. (2002).
The coordinates of Oph-N9 are taken from Persson et al. (1998).
}
\begin{tabular}{@{}lcccccccc@{}}
\hline
      Object   &  RA offset & Dec. offset &   RA  &  Dec.      &  $P$    &  $\theta$  & Mag$_{2.0 \mu m}$  & Mag$_{2.0 \mu m}$ \\
               &  (arcsec)  &  (arcsec)   & (J2000) & (J2000)  & (percent)  &  (degrees) & 2006 Aug 31 & 2006 Oct 8 \\
\hline
Stars in Mon R2 &&&&&&&& \\
Star O (IRS2) &  -18.61 & 3.82 & 06:07:46.59 &  $-$06:22:52.5 &   $7 \pm 1$ &  $167 \pm 3$ & -  & 14.91 \\
Star P (IRS2) &  -17.21 & 1.36 & 06:07:46.68 &  $-$06:22:54.9 &   $5 \pm 1$ &  $177 \pm 3$ & - & 14.62 \\
Star Q (IRS2) &  -16.71 & -1.30 & 06:07:46.71 & $-$06:22:57.6 &   $8 \pm 1$ &  $171  \pm 7$ & - & 14.97 \\   
Star G &    -6.80 &   2.28 &    06:07:47.38 &  $-$06:22:54.0 &   $11 \pm 1$ &  $173 \pm 2$  & 14.69 & 15.07  \\
Star H &    -4.58 &  -0.38 &    06:07:47.53 &  $-$06:22:56.7 &   $3 \pm 1$ &   $8 \pm  6$  & 14.41 & 14.44  \\
Star I &    -2.60 &  -4.95 &    06:07:47.66 &  $-$06:22:61.2 &     - &  -  & 16.39 & 16.35 \\
Star N &    -2.56 &   3.11 &    06:07:47.66 &  $-$06:22:53.2 &      - &   -  & 17.60 & 17.54 \\
Star E &    -2.38 &   1.14 &    06:07:47.68 &  $-$06:22:55.1 &     $ 1 \pm 1$ &  $ 20 \pm 20$  & 12.99 & 13.16 \\
Star F &    -1.48 &   0.02 &    06:07:47.74 &  $-$06:22:56.3 &      -   & - & 14.47 &  14.29 \\
Star J &    -0.08 &  -7.53 &    06:07:47.83 &  $-$06:22:63.8 &    $  2 \pm 1$ &  $ 15 \pm 9$  & 12.73 & 12.74 \\
Star A &     0.00 &   0.00 &    06:07:47.84 &  $-$06:22:56.3 &    $  9 \pm 1$ &  $112 \pm 2$  & 9.25 &  9.21 \\
Star B &     0.30 &   0.93 &    06:07:47.86 &  $-$06:22:55.4 &    $ 12 \pm 1$ &  $166 \pm 1$  & 9.46 &  9.49 \\
Star C &     0.65 &   0.99 &    06:07:47.88 &  $-$06:22:55.3 &    $  7 \pm 1$ &  $174 \pm 2$  & 10.17 & 10.30 \\
Star D &     0.77 &   2.09 &    06:07:47.89 &  $-$06:22:54.2 &    $  5 \pm 1$ &  $147 \pm 4$  & 12.56 & 12.51 \\
Star K &     3.35 &  -5.68 &    06:07:48.06 &  $-$06:22:62.0 &      - &  -  & 14.96 & 14.98 \\
Star M &     4.34 &   6.36 &    06:07:48.13 &  $-$06:22:49.9 &    $ 11 \pm 3$ &  $121 \pm 7$  & 17.18 & 17.03 \\
Star L &    11.76 &   0.72 &    06:07:48.62 &  $-$06:22:55.6 &      - &  -  &  15.16 & - \\
 &&&&&&&& \\
Oph-N9 &&&&&&& 2006 Apr 10 & \\
Oph-N9   &   -   &   - &  16:27:13.3 &  $-$24:41:34  &  $8.6\pm  1$  &   $ 32 \pm 2$  &    10.26  &\\
\hline
\end{tabular}
\end{minipage}
\end{table*}



\setcounter{table}{2}
\begin{table*}
\centering
\begin{minipage}{160mm}
\caption{Stellar polarization and photometry measurements of S140-IRS1.
Polarization uncertainties are described in Table~2.
The absolute coordinates are derived from the 2MASS detection of S140-IRS1 (2MASS 22191827+6318458),
the only well-detected 2MASS source in the field of view.
}
\begin{tabular}{@{}lcccccccc@{}}
\hline
      Object   &  RA offset & Dec. offset &   RA  &  Dec.      &  $P$    &  $\theta$  & Mag$_{2.0 \mu m}$  & Mag$_{2.0 \mu m}$ \\
               &  (arcsec)  &  (arcsec)   & (J2000) & (J2000)  & (percent)  &  (degrees) & 2006 Mar 30 & 2006 Apr 22 \\
\hline
       1  &  $-$14.11 &   6.92  &   22:19:16.18 &  63:18:52.7   &    - & -  &  16.21 & 16.30  \\ 
       2  &   $-$8.32 &  10.49  &   22:19:17.04 &  63:18:56.3   &    - & -  &  18.75 & 19.21  \\ 
       3  &   $-$8.12 &  10.86  &   22:19:17.07 &  63:18:56.7   &    - & -  &  18.02 & 18.82  \\ 
       4  &    $-$6.67 &   7.94 &   22:19:17.29 &  63:18:53.8  &     $5 \pm 1$  &  $13 \pm 1$  &  13.20 & 13.31  \\ 
       5  &  $-$5.17 &  11.35   &   22:19:17.51 &  63:18:57.2    &   - &  -  &  15.70 & 15.70 \\ 
       6  &   $-$2.05 &  $-$3.55  & 22:19:17.97 &  63:18:42.3   &    $3 \pm 1$ &  $127 \pm 5$  &  12.21 & 12.35  \\ 
       7  &  $-$0.50  &  5.41   &   22:19:18.20 &  63:18:51.2    &   - &  -  &  15.40 & 15.50  \\ 
       8 S140-IRS1  &     0.00 &   0.00 &    22:19:18.28 &  63:18:45.8  &     $4 \pm 1$  &  $40 \pm 3$   &  7.72 & 7.67  \\ 
       9  &    0.19  &  4.04  &    22:19:18.70 &  63:18:49.9   &    - & -  &  15.78 & 15.76  \\ 
      10  &     1.60 &   7.36 &    22:19:18.31 &  63:18:53.2  &     $3 \pm 1$  &  $45 \pm 6$  &  12.25 & 12.16  \\ 
      11  &    6.32  & $-$3.20  &  22:19:19.21 &  63:18:42.6   &    $5 \pm 1$  &  $94 \pm 5$  &  15.89 & 16.01  \\ 
      12  &    6.82  &  8.48  &    22:19:19.29 &  63:18:54.3   &    - &  -  &  16.95 & 16.82  \\ 
      13  &     7.71 &   0.78 &    22:19:19.42 &  63:18:46.6  &    $16 \pm 1$ &  $100 \pm 4$  &  15.03 & 15.17  \\ 
\hline
\end{tabular}
\end{minipage}
\end{table*}



\setcounter{table}{3}
\begin{table*}
\centering
\begin{minipage}{160mm}
\caption{Stellar polarization and photometry measurements of AFGL 2591.
Polarization uncertainties are described in Table~2.
The absolute coordinates are derived from the 2MASS detection of AFGL~2591 (2MASS 20292486+4011194),
the only well-detected 2MASS source in the field of view.
}
\begin{tabular}{@{}lcccccccc@{}}
\hline
      Object   &  RA offset & Dec. offset &   RA  &  Dec.      &  $P$    &  $\theta$  & Mag$_{2.0 \mu m}$  & Mag$_{2.0 \mu m}$ \\
               &  (arcsec)  &  (arcsec)   & (J2000) & (J2000)  & (percent)  &  (degrees) & 2006 May 17 & 2006 July 1 \\
\hline
       1 &    $-$10.44 & $-$10.17 &    20:29:23.96 &  40:11:09.2  &  $2 \pm  1$  & $ 71 \pm  9$  &  12.49 & - \\ 
       2 &     $-$7.74 &   0.51 &    20:29:24.19 &  40:11:19.9  &     $2 \pm  1$  & $ 63 \pm  9$  & 14.56 & 14.50 \\ 
       3 &     $-$7.73 &   2.01 &    20:29:24.19 &  40:11:21.4  &    - &   -  & 15.17 &  14.96 \\ 
       4 &     $-$7.10 &  $-$7.80 &    20:29:24.25 &  40:11:11.6  &  - &    -  & 15.10 & 15.34 \\ 
       5 &     $-$6.61 &  $-$6.98 &    20:29:24.29 &  40:11:12.4  &  - &  -  & 16.84 & 16.81 \\ 
       6 &     $-$6.44 &   8.46 &    20:29:24.31 &  40:11:27.9  &    $16 \pm  1$ &  $125 \pm  1$  & 15.15 & - \\ 
       7 &     $-$5.12 &  $-$8.74 &    20:29:24.42 &  40:11:10.7  &   - &   -  & 17.62 & 17.58 \\ 
       8 &     $-$4.57 &   9.99 &    20:29:24.47 &  40:11:29.4  &   - &   -  & 18.17 & - \\ 
       9 &     $-$4.28 &   0.58 &    20:29:24.49 &  40:11:20.0  &    $ 8 \pm  1$ &  $ 98 \pm 2$  & 14.28 & 14.27 \\ 
      10 &     $-$4.20 & $-$10.43 &    20:29:24.50 &  40:11:09.0  &   - &  -  & 15.57 & 15.56 \\ 
      11 &     $-$4.02 &  $-$7.74 &    20:29:24.52 &  40:11:11.7  &   - &  -  & 16.70 & 17.11 \\ 
      12 &     $-$3.84 &   0.31 &    20:29:24.53 &  40:11:19.7  &    $14 \pm  1$ &  $104 \pm  2$  & 14.50 & 14.43 \\ 
      13 &     $-$3.37 &  $-$0.87 &    20:29:24.57 &  40:11:18.5  &  - &  -  & 16.24 & 16.08 \\ 
      14 &     $-$2.91 & $-$15.04 &    20:29:24.61 &  40:11:04.4  &   - & -  & 16.07 & 16.05 \\ 
      15 &     $-$2.52 & $-$13.24 &    20:29:24.65 &  40:11:06.2  &   - &  -  & 16.56 & 16.61 \\ 
      16 &     $-$2.50 &  $-$0.89 &    20:29:24.65 &  40:11:18.5  &  - &  -  & 15.22 & 15.29 \\ 
      17 &     $-$1.95 &   3.32 &    20:29:24.70 &  40:11:22.7  &   - &  -  & 17.16 & 16.92 \\ 
      18 &     $-$1.55 &  $-$0.36 &    20:29:24.73 &  40:11:19.1  & $6 \pm 1$   & $133 \pm 4$   & 15.32 & 15.35 \\ 
      19 &     $-$1.53 &   7.40 &    20:29:24.73 &  40:11:26.8  &   - &  -  & 16.54 & 16.62 \\ 
      20 &     $-$1.00 &   4.23 &    20:29:24.78 &  40:11:23.6  &   - &  -  & 15.60 & 15.59 \\ 
      21 &     $-$0.67 &   9.98 &    20:29:24.81 &  40:11:29.4  &   - &  -  & 15.55 & 15.53 \\ 
      22 &     $-$0.29 &  $-$8.86 &    20:29:24.84 &  40:11:10.6  &  $   8 \pm  2$ &  $165 \pm 11$  & 17.07 & 17.03 \\ 
      23 &     $-$0.20 &   2.89 &    20:29:24.85 &  40:11:22.3  &    -  &  -  & 15.50 & 15.54 \\ 
      24 AFGL~2591 &     0.00  &  0.00  & 20:29:24.87 &  40:11:19.4  &  $  16 \pm  2$ & $ 171 \pm  2$  & 7.53  & 7.57 \\ 
      25 &     0.59  &  6.41  &   20:29:24.92  & 40:11:25.8   &  -  &  -  & 17.04 & 16.93 \\ 
      26 &      1.49 & $-$12.46 &    20:29:25.00 &  40:11:07.0  &   - &  - & 17.77 &  17.81 \\ 
      27 &      2.14 &  $-$7.66 &    20:29:25.05 &  40:11:11.8  &  - &  - & 18.13 &  18.00 \\ 
      28 &      2.40 & $-$13.66 &    20:29:25.08 &  40:11:05.7  &   - &  - & - &  16.75 \\ 
      29 &      2.74 &  $-$8.85 &    20:29:25.11 &  40:11:10.6  &  - &  - & 18.44 &  18.30 \\ 
      30 &      3.58 &   8.03 &    20:29:25.18 &  40:11:27.4  &  - &  - & 18.24 &  18.33 \\ 
      31 &      4.22 &  $-$9.89 &    20:29:25.24 &  40:11:09.5  &  - &   - & 16.49 &  16.74 \\ 
      32 &      4.48 &  $-$5.03 &    20:29:25.26 &  40:11:14.4  &  $5 \pm 2$ & $167 \pm 7$ & 15.49 &  15.47 \\ 
      33 &      4.70 &  $-$0.60 &    20:29:25.28 &  40:11:18.8  &   - &  - & 17.76 &  18.01 \\ 
      34 &      5.25 &   8.18 &    20:29:25.33 &  40:11:27.6  &  - &  - & 16.24 &  16.21 \\ 
      35  &     5.32 &   0.56 &    20:29:25.33 &  40:11:20.0  &   $ 13 \pm  3 $ &  $42 \pm  7$ & 17.19 &  16.75 \\ 
      36  &     6.88 & $-$10.27 &    20:29:25.47 &  40:11:09.1  &  $   8 \pm  2 $ & $151 \pm 13$ & 16.66 &  16.79 \\ 
      37  &     9.09 &  $-$0.12 &    20:29:25.66 &  40:11:19.3  &   - &  - & 17.09 &  17.06 \\ 
\hline
\end{tabular}
\end{minipage}
\end{table*}


The positions and fluxes were measured for each detected star 
in each of the three POL filters. 
The positions of the stars were measured by fitting Gaussian functions to the 
cores of the star images and 
the fluxes were measured by aperture photometry of each star on 
the median-combined but unrotated POL images.
The polarization parameters 
 $I$, $Q$, $U$, $P$, and $\theta$ were computed from the measured fluxes 
by multiplying the flux vector by the same matrix that was used to 
compute $I$, $Q$, and $U$ from the combined dithered images
(Batcheldor et al. 2006, 2009). 
The star measurements used a circular aperture with a radius of 2.5 pixels 
(to the minimum of the first Airy dark ring) and the background was measured in 
a ring of radii 5 to 7 pixels (just outside the first Airy bright ring).
This aperture is too small for accurate measurement of isolated polarization standard stars,
as was noted by Batcheldor et al. (2006), 
who estimated the accuracy of measurements of $P$ and $\theta$ 
from simulated star images with sub-pixel misalignment.
However, the large and polarized background in our images 
also introduces substantial uncertainty if larger apertures are used.
Consequently, except for the very bright central stars of AFGL 2591 and S140-IRS1, 
we preferred the small aperture but estimate that there could 
be an additional systematic uncertainty of  
a few tenths of a percent to the measured percentage polarization. 

In fact, the difference between the visits is sometimes larger than a percent, 
in which case the uncertainty in the tables is equal to 0.5 times the difference 
of the measured values with the minimum uncertainty given as 1 percent
to account for systematic effects. 
Simpson et al. (2009) describe how the uncertainties in the polarization measurements 
were estimated from first principles.

Stellar magnitudes were estimated from the Stokes $I$ flux for each star for each visit.
The NICMOS calibration\footnote{http://www.stsci.edu/hst/nicmos/performance/photometry}
assumes an aperture radius of 0.5 arcsec. 
However, we found that except for S140-IRS1, AFGL 2591 YSO, and six other stars much brighter
than the background, the background fluxes subtracted from the measured stars plus background 
are larger to much larger than the final star fluxes if an aperture radius as large 
as 0.5 arcsec is used. 
Consequently, for estimates of the magnitudes 
we used the larger 0.5 arcsec radius aperture for these brighter stars but 
for all other stars we used the smaller 0.19-arcsec radius aperture (2.5 pixels) 
that was used for measuring the polarization in order to minimize 
the uncertainty due to background subtraction. 
These measured fluxes were then increased to 
those expected for the 0.5 arcsec calibration radius 
by multiplying by the corresponding ratio of synthetic PSFs computed 
with {\it HST} Tiny Tim (Krist, Hook, \& Stoehr 2011) for a colour temperature of 500~K,
appropriate for YSOs. 

Consequently, the sources of uncertainty in the photometry include,
in addition to background subtraction and the usual photon noise for faint sources,  
uncertainties in the PSF computation, 
the fact that the coronagraph images were not dithered 
(as is assumed for the NICMOS flux calibration), 
and the fact that the coronagraph uses a slightly different focus position 
(not modelled by Tiny Tim) from that used by normal images.
No attempt was made to quantify the uncertainties in the photometry; 
however, we estimate that they are of the order of 5 -- 10 percent 
based on the reproducibility of the stellar magnitudes measured 
from images obtained in the different visits and the different types of observation 
(coronagraph and non-coronagraph). 
Although most stars appear to be $\sim$ constant from one visit to the next, 
a few have larger differences in the fluxes than would be expected from the errors. 
The two objects whose time variability was confirmed by blinking the images are 
Star G in Mon R2-IRS3 and Star 11 in AFGL 2591. 
There is an additional uncertainty in the flux of 5 - 10 percent 
due to the calibration uncertainty that does not affect any estimates of the variability.
Additional discussion of stellar fluxes computed from NICMOS polarization observations
is given by Simpson et al. (2006).

The polarizations and uncertainties are given in Tables 2 -- 4
and the locations of the stars are plotted in Figs. 1(c), 2(c), 3(c), and 4.
With the exception of Star 32 in AFGL 2591, 
values for polarization are listed only for stars 
that are detected on the $Q$ and/or $U$ images 
and have $P/\sigma_P > 4$
(AFGL 2591 Star 32 has good signal/noise for the PA, although 
its polarization is less certain).
Because of the possibility of systematic uncertainties, the minimum 
tabulated $\sigma_P$ is 1 percent; however, the stars in the tables 
with $P = 1 \pm 1$ or $2 \pm 1$ are real detections.
These are relatively bright stars and the computed uncertainties in $P$ 
are $\la 0.1 - 0.3$ percent.

Relative positions (offsets) for all but the faintest stars are accurate to $\sim 0.02$ arcsec.
Absolute positions used offsets from the Two Micron All Sky Survey 
(2MASS, Skrutskie et al. 2006) positions of AFGL 2591 and S140-IRS1  
and the position of Mon R2-IRS3 Star A from Preibisch et al. (2002).

For Mon R2 we use the designations of the five brightest stars in IRS3 
from Preibisch et al. (2002), 
taken in order of $K$-band brightness from their NIR speckle imaging.
As a result, the star names are letters instead of numbers. 
We note that Stars O, P, and Q are located in IRS2, not IRS3.

\section{Results}

Our data show that all four YSOs (two YSOs in Mon R2-IRS3) appear as monopolar outflows 
with highly polarized diffuse emission (Figs. 1 -- 3).
The illuminating stars of all three sources are also visible; 
we infer that these are the YSOs that are the sources of the outflows. 
They are identified by the centrosymmetric polarization vectors in the diffuse emission.
Moreover, the emission from the YSOs themselves is also significantly polarized 
with polarization PAs, $\theta$, approximately perpendicular to the outflow directions.
We will discuss each of the sources in turn, and then compare their polarized images 
to Monte Carlo scattering models. 

\subsection{Mon R2-IRS3}

\begin{figure}
\includegraphics[width=84mm]{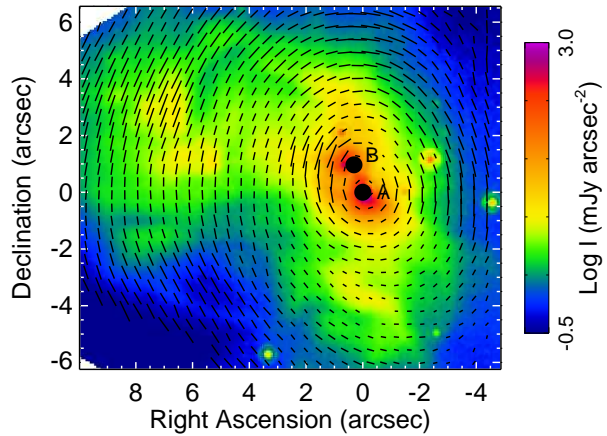}
\caption{
Enlargement of the central region of Mon R2-IRS3.
Black dots replace the stars A and B: Star A  
because it was placed behind the coronagraph hole 
and Star B because it and its polarized PSF were subtracted from the image.
}
\end{figure}

Figs. 1 and 4 show the {\it HST} images of Mon R2-IRS3 with
Fig. 1 including the eastern edge of IRS2. 
We see that IRS3 consists of a compact cluster of YSOs.
The two brightest YSOs, named Star A and Star B by Preibisch et al. (2002), 
show monopolar outflows at approximately orthogonal directions
(Star A's outflow extends to the south and Star B's outflow extends to the east). 
In Fig. 5 we show an enlargement of these orthogonal outflows, 
where we have subtracted the polarized PSF 
(derived from measurements of Oph-N9 centred in the Camera 2 array) 
from Star B as well as Star A. 
From the alignment of the polarization vectors, we infer that the east outflow 
is illuminated by Star B and the south outflow by Star A. 

Star B is significantly redder than the other stars in IRS3. 
This is seen in both the colour version of Fig. 4 and in the values of $H-K$ 
measured by Preibisch et al. (2002), where $H-K$ equals 4.4 for Star B 
and 2.9 (Star A) or bluer for the other five stars in their table of stellar magnitudes. 
We suggest that the red colour of Star B is due to extinction 
and that some of the extinction of Star B is due to 
it being located behind the YSO envelope of Star A.

At least some of the other stars in this grouping are probably members of 
the same cluster as Stars A and B. 
Preibisch et al. (2002) find that Stars E, C, and A are X-ray sources 
with typical YSO X-ray properties.
Assuming that they are all at the same distance (830 pc) and using their NICMOS photometry, 
Preibisch et al. (2002) estimate that the masses of Stars A, B, C, D, and E are 
in the ranges 12 -- 15, 8 -- 12, 5 -- 10, 2 -- 5, and $\sim 1$ M$_\odot$, respectively.

Alvarez, Hoare, \& Lucas (2004b) modelled the south lobe of Star A 
with a Monte Carlo radiation transfer code. 
They assume for their models that the outflow axis is tipped towards the earth by $\sim 45 ^\circ$.
With this inclination angle, at least part of the YSO envelope should be obscuring the star.
In fact, Star A is present as a {\it HST} point source 
(i.e., exhibiting the first Airy bright ring of the {\it HST} PSF) only at 2.07 \micron; 
at shorter wavelengths the stellar point source is obscured 
by the bright scattered light of the south scattering lobe.
We conclude that an inclination of $\sim 45 ^\circ$ is not unreasonable for the Star A outflow.

\subsection{S140-IRS1}

\begin{figure}
\includegraphics[width=84mm]{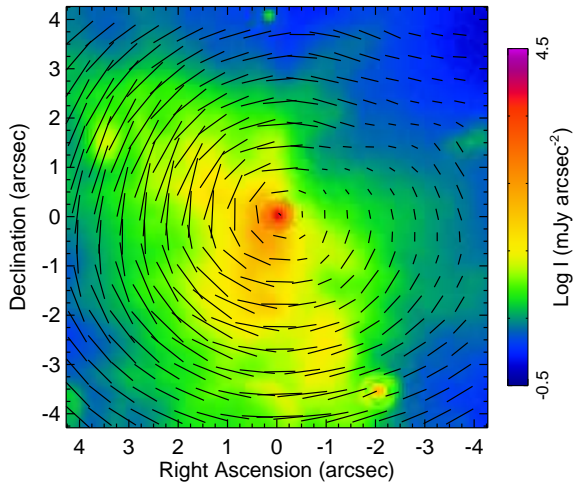}
\caption{
Enlargement of the central region of S140-IRS1.
}
\end{figure}

The image of S140-IRS1 in 
Fig. 2 shows a strong central source surrounded by highly polarized nebulosity. 
The polarization vectors are generally centrally symmetric about IRS1 in the regions of 
high polarization, with the exception that a region to the east of IRS1 has its own 
centrally symmetric polarization vectors.
These indicate the location of IRS3, which was just off the edge of the NICMOS array.

The nebulosity is very clumpy, giving the appearance of a bipolar outflow to the 
northeast and southwest, along with nebulosity in the southeast.
The nebulosity to the southeast has a concentration at PA $\sim 154$\degr,
which can be seen in the enlargement of the region shown in Fig. 6. 
This was first observed at 2.165 \micron\ by Schertl et al. (2000),
who describe the concentration as the scattered light off the inner surface 
of the evacuated cavity of the bipolar outflow seen in CO by Minchin et al. (1993).
Schertl et al. (2000) speculate that the opposite northwest lobe is obscured 
by the optically thick envelope or disc of IRS1. 
We agree that this is a likely explanation since our substantially more sensitive 
NICMOS observations also do not detect any northwest outflow.

The PA of the polarization vector for IRS1, $\theta = 40 \pm 3$\degr, 
is very close to the PA
($\sim 44$\degr) of the elongated 43 GHz radio source observed by Hoare (2006), 
which he attributes to an equatorial wind from the surface of a disc
(see Gibb \& Hoare 2007 for further discussion of the radio source).
The symmetry of the region shown in Fig.~6 is such that 
any outflow could indeed be oriented with a PA on the order of 135\degr.
If so, the polarization PA for IRS1 is perpendicular to the outflow 
and parallel to the suggested disc of Hoare (2006). 

However, the additional curved features seen to the northeast in Fig. 2(b) 
would not be part of this outflow structure 
if the outflow has a PA of 135\degr. 
Minchin, Ward-Thompson \& White (1995) detected 
additional very cold sources to the southwest and northwest.
Maud et al. (2013) suggest that the curved features seen to the northeast in Fig. 2(b)
are shocks from an outflow originating in the southwest cold source, SMM1. 
Shocked H$_2$ was also seen at this northeast position by Preibisch \& Smith (2002). 
If so, this shocked outflow must have broken through the S140-IRS1 envelope 
in the direction towards the earth because the nebulosity is clearly 
illuminated by IRS1. 
It must be located close to the plane of the sky that includes IRS1 because 
the polarization is so high, $\sim 83$ percent, in these features. 

\subsection{AFGL 2591}

\begin{figure*}
\includegraphics[width=174mm]{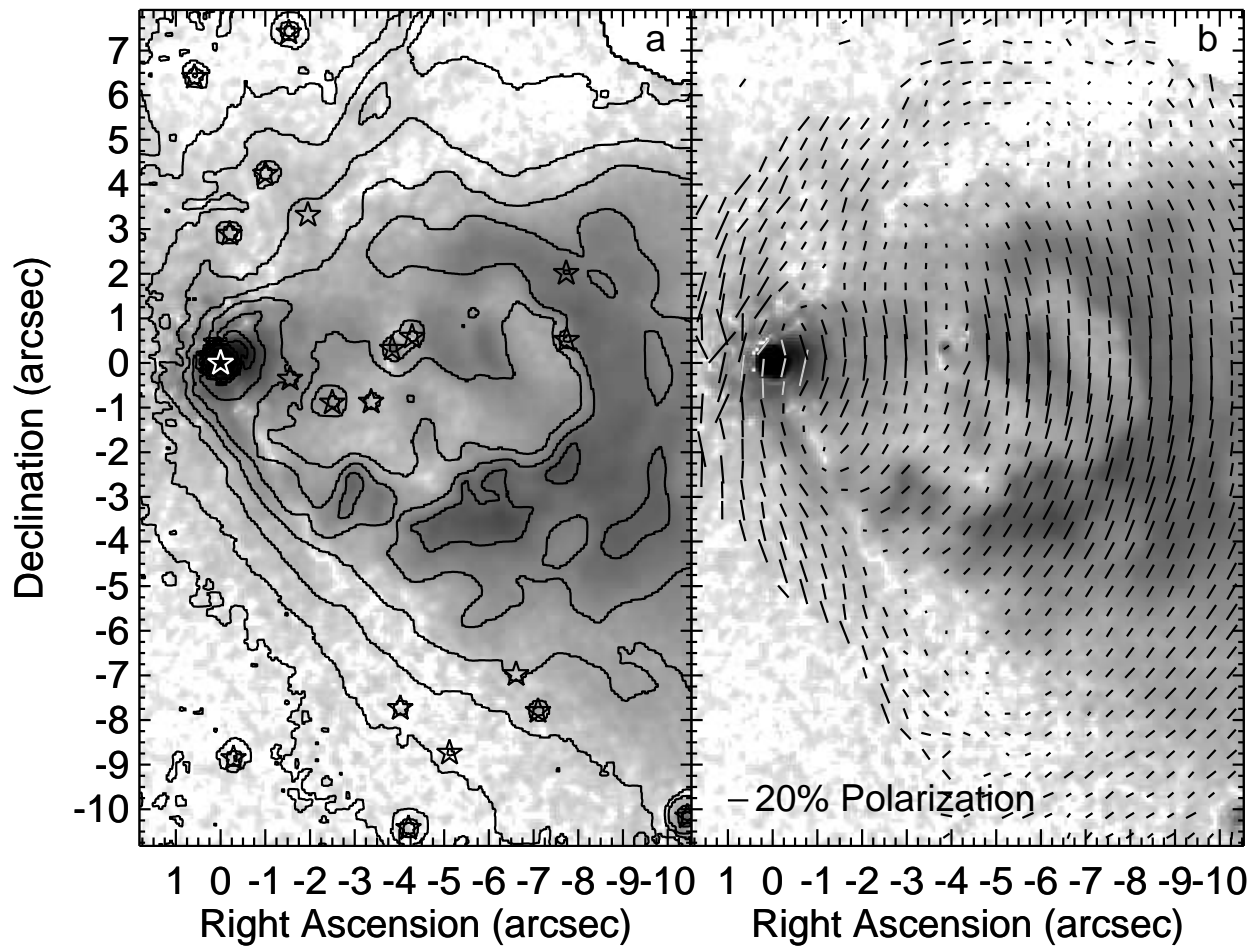}
\caption{
Enlargement of the central region of AFGL 2591. 
Panel (a): gray-scale image of the logarithm of the polarized intensity. 
The contours show the total intensity in units of 0.1 to 100 mJy arcsec$^{-2}$
with steps of factors of $10^{1/3}$. 
The observed stars from Table~4 are marked.
Panel (b): gray-scale image of the logarithm of the polarized intensity.  
The polarization vectors are plotted in black or gray.
}
\end{figure*}

The image of AFGL 2591 in Fig. 3 shows a monopolar outflow 
with several loops of up to 40 percent polarized, scattered light 
about 10~arcsec to the west of the bright YSO.
At distances larger than about 5 -- 6~arcsec, the perpendiculars to the polarization vectors
all point towards the YSO, as would be expected for single scattering.
However, at closer distances, especially in the central several arcsec, 
the polarization vectors in those positions along the 
limb of the scattered light region of the outflow mostly lie in a line 
perpendicular to the outflow direction.
That is, their PAs are similar to that of 
the AFGL 2591 YSO itself: 172\degr.
This is shown more clearly in Fig.~7, which is an enlargement 
of the central $\pm 10$~arcsec. 
Beyond $\sim 3$ arcsec from the YSO and continuing along the limb, 
the polarization vectors change their PAs to remain somewhat parallel to the limb 
with the perpendiculars to these vectors pointing into the centre of the outflow 
and not back to the bright YSO. 
This is especially apparent at the top of Fig.~7b  
(RA $\sim -4$ to $-8$ arcsec and Dec $\sim +7$ arcsec) 
and near the bottom of the figure 
(RA $\sim -4$ to $-8$ arcsec and Dec $\sim -9$ arcsec). 

The gray-scale of Fig. 7 is the polarized flux,  $I_P = I \times P$.
One of the most striking aspects of Fig.~7(a) is that $I_P$ is 
discordant with the contours representing the total intensity, $I$, 
which is all scattered light (see Fig. 3b). 
Those regions where the polarization vectors change directions emit very little 
polarized intensity
(compare the polarization vectors and the gray-scale in Fig. 7b), 
even though their total intensity (Fig. 7a) is still substantial.

There are three possible reasons for having scattered light with little polarization: 
(1) The light is produced by forward or backward scattering by spherical particles.
(2) The particles are elongated and are illuminated edge-on with scattering angles 
at intermediate angles such as $\sim 45$\degr\ or 135\degr\ 
(Matsumura \& Seki 1996; Whitney \& Wolff 2002; Wolf, Voshchinnikov \& Henning 2002).
(3) The dust consists of aligned grains, which are partially optically thick 
and the polarization angle of the dust in the background 
is rotated by 90\degr\ by the absorbing dust in the foreground 
(Whitney \& Wolff 2002). 

In the next section we compare the data to models using both spherical and 
elongated, aligned grains. 
We will show that the models with spherical grains produce too little polarization 
compared to the observations; the likely reason is that the scattering is 
mostly forward scattering at the low inclination angles of the best fitting models. 
Better agreement with the observations occurs for models with aligned grains.
This is true for both the high polarization seen at the location of the illuminating star 
and at large distances along the outflow axis, and for the regions of low 
polarized intensity described above.

\section{Comparison to models}


\setcounter{table}{4}
\begin{table*}
\centering
\begin{minipage}{170mm}
\caption{Parameters of fitted models.
For each YSO the two top models are tabulated because they demonstrate the parameter range of acceptable fits.
The first group of parameters are copied from the input files that were used by 
Robitaille et al. (2006)$^3$.
Additional parameters not listed here are either irrelevant 
or are the same for all models 
(such as the envelope parameters CSHAPE = 'POLYN', EX1=1.5 for the cavity shape exponent, 
Z01=0.0, and EXF=0.0, which are cavity wall and cavity density parameters).
All models in this paper were computed using $10^8$ photons.
}
\begin{tabular}{@{}lccccl@{}}
\hline
\multicolumn{1}{l}{Parameter name} & \multicolumn{2}{c}{S140-IRS1} & \multicolumn{2}{c}{AFGL 2591} & Description \\
               & Model 1 & Model 2 & Model 1 & Model 2 & \\
\hline
\multicolumn{4}{l}{Parameters from the two best-fitting SED models}\\
SED model ID  &   3000319 & 3004583 &  3007097  & 3018960 &  Model number from Robitaille et al. (2006) \\
RSTAR         &   190.170 & 185.080 &  13.259   & 12.135  &  Stellar radius in Solar radii \\
TSTAR         &   4189.7  & 4212.9  &  36299.0  & 36002.0 &  Blackbody temperature of central star (K) \\
MASSC         &   19.470  & 19.242  &  37.568   & 34.012  &  Mass of central star (in Solar masses) \\
MASSD         &    0.0    & 0.0     &   0.0     & 0.1068  &  Disc mass in Solar masses \\
RMAXD      &      -  & -  & -  & 74.448 &  Maximum disc radius in au \\
RMAX      &     2.793E+04 & 9.939E+04 & 1.00E+05 & 1.00E+05 &  Maximum envelope radius in au \\
RATE      &     4.384E$-$04 & 2.349E$-$04 & 4.498E$-$03 & 1.344E$-$03 & Envelope mass infall rate (solar masses year$^{-1}$) \\
RC        &      3.223   &  5.099   & 8.901     &  74.448 & Envelope centrifugal radius (au) \\
THET1     &      5.038   & 2.296    &  7.971     & 5.966 &  Opening angle of cavity wall (degrees) \\
RHOCONST1 &    4.443E$-$20 & 6.391E$-$20 & 4.367E$-$20  & 2.548E$-$20 &  Coefficient for cavity density distribution \\
RHOAMB    &    3.767E$-$21 & 1.207E$-$20 & 1.670e$-$20  & 8.381E$-$21 &  Ambient density (gm cm$^{-3}$) \\

&&&&& \\
\multicolumn{1}{l}{Output results} &&&&& \\
L/L$_\odot$ &   1.00E+04 & 9.69E+03  & 2.74E+05 & 2.24E+05  & Luminosity (in Solar luminosities) \\
M$_{\rm env}$/M$_\odot$& 34.6 &   131    &   1720      & 543 & Envelope mass (in Solar masses)  \\

&&&&& \\
\multicolumn{4}{l}{Modified parameters from the {\sc ttsscat} and {\sc ttsscat\_al} models}  \\

THETE      &     12.0    & 20.0 &  15.0  & 15.0     &  Inclination angle (degrees) \\
THET1      &     5.038   & 4.3 &  10.9  & 7.5     &  Opening angle of cavity wall (degrees) \\
RHOCONST1  &   4.443E$-$20 & 6.391E$-$20 & 4.367E$-$22 & 2.548E$-$22   &  Coefficient for cavity density distribution \\
RHOAMB     &   3.767E$-$21 & 1.207E$-$20 & 1.670e$-$22 & 8.381E$-$23  &  Ambient density (gm cm$^{-3}$) \\

&&&&& \\
\multicolumn{4}{l}{Model magnetic field parameters (see text for descriptions)} \\
$H_Z$       &    268.6  & 115.9 &   8901 & 2481.6     &   Scale height for decrease of $B_\phi$ with $Z$ in au \\
$C_\phi$   &      2.0E+06 & 1.0E+6 &  3.0E+10 & 6.26E+08 &   Unitless constant in the equation for $B_\phi$ \\

POL-YSO    &    0.04  & 0.04 &  0.90  & 0.73      &  Resulting polarization of the illuminating YSO \\
Z1         &    1800  & 1000 &   $\sim$ 1.E+05 & $\sim 34000$   &  Height in au from YSO to where $B_\phi/B_Z=1$   \\
\hline
\end{tabular}
\end{minipage}
\end{table*}


\begin{figure}
\includegraphics[width=84mm]{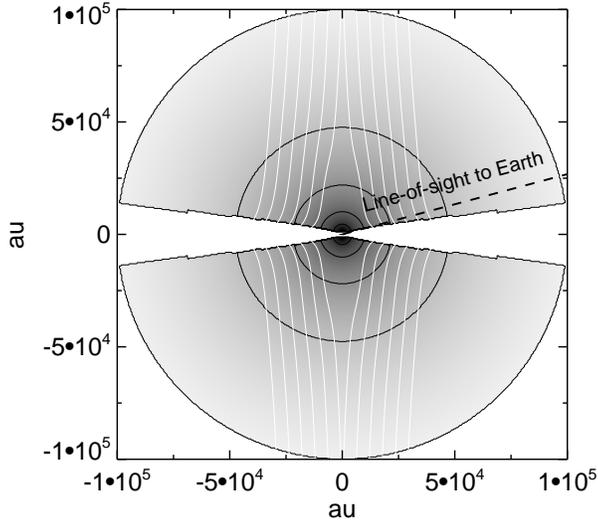}
\caption{
Cross-section through AFGL 2591 model 3018960 
showing the geometry of the outflow cavity 
and the 15\degr\ inclination angle towards the Earth. 
The gray-scale represents the log of the model density;
the polar axes lie to the right and left and 
are surrounded by low-density cavities with cavity opening angles of 7.5\degr. 
The black contours mark the locations where the density equals 
$10^{-7.5}$ to $10^{-5.5}$ gm cm$^{-3}$ with steps of factors of $10^{0.5}$.
The white contours are an example of the magnetic field components 
described by equations (1) -- (3); they represent the locations where the ratio 
$B_\phi/B_Z$ equals 0.1, 1, 10, $10^2$, $10^3$, and $10^4$. 
}
\end{figure}

To further analyse the polarization images of S140-IRS1 and AFGL 2591, 
we have modelled the data using 
the Monte Carlo scattering routine 
{\sc ttsscat}\footnote{HO-CHUNK.ttsscat.20090521 is available from http://gemelli.colorado.edu/$\sim$bwhitney/codes} 
of Whitney \& Hartmann (1992, 1993).
The models consist of an envelope, somewhat flattened by rotation, an outflow cavity, 
and an optional circumstellar disc. 
A cross-section through a model showing the direction of the line of sight 
is illustrated in Fig. 8. 
The mass of the envelope is determined by the accretion rate; 
for the massive YSOs in this study, the accretion rate is so high that the 
flattened part of the envelope near the equatorial plane can be thought of 
as a dense, optically thick toroid, such that any Keplerian disc is not significant.
Additional details of the physics of these models can be found in the papers 
by Whitney \& Hartmann (1993), Stark et al. (2006), and Robitaille et al. (2006). 
Stark et al. (2006) also give a number of examples of models computed with this program.
As in that paper, we assume the same `curved' cavity shape 
and that the dust grains are spherical 
with parameters for the dust size and composition given by Kim, Martin, \& Hendry (1994). 
Other cavity shapes and dust parameters were tried but gave poorer matches to the observations.
The models in this paper were all computed with grain scattering parameters 
for a wavelength of 2.2 \micron, only slightly longer than the NICMOS POL filters' 2.0 \micron.

Our criteria for a good fit are that the model spectral energy distribution (SED)  
is a good approximation to the observed SED, 
that the relative brightness of the illuminating star compared to the scattered light intensity 
approximates the data, 
and that the general shape of the model scattered intensity approximates the shape of the 
observed outflow. 
Fitting the first criterion requires that the model have approximately the correct luminosity, 
distance, and envelope mass. 
Fitting the second criterion requires an estimation of both the model inclination angle 
and the cavity opening angle in order to have a significant, but not too large, optical depth 
along the line of sight to the star due to the dust in the model's extended envelope, 
where the envelope parameters are determined by the first criterion. 
The last criterion should be much more important; however, both S140-IRS1 and AFGL 2591 
appear very clumpy and possibly affected by secondary sources in their vicinities. 
These will be discussed along with the models of each source. 

For a first guess for the input parameters, we start by using the best fits from  
the online SED fitting program of Robitaille et al. 
(2006, 2007)\footnote{http://caravan.astro.wisc.edu/protostars} 
and infrared fluxes from the literature (Figs. 9 and 10).
Because most of the code is the same, the parameters from the SED-fitting program 
(Robitaille et al. 2007)
can be directly entered into {\sc ttsscat}. 
Table 5 contains a description of these parameters. 
We note that when the table says the disc mass is zero, 
it means that the mass of any {\it dusty} disc is zero because any dust 
within the normal disc radius of a few tens of au would be destroyed 
by the extremely luminous central protostars. 
Gaseous discs were not included in the Robitaille et al. (2006) models 
because they do not appreciably change the SED. 
As a test case, a radius 500-au dusty disc was added to several models; 
no difference was seen in the computed 2 \micron\ scattered light and polarization
because of the very high optical depth towards the centre.

We do not compute scattering models for Mon R2-IRS3 because 
(1) it has at least 2 YSOs of similar 2-\micron\ brightness and hence a SED for 
a single YSO could not be found,  
and (2) comparable models were described by Alvarez et al. (2004b) for its Star A.

\begin{figure}
\includegraphics[width=84mm]{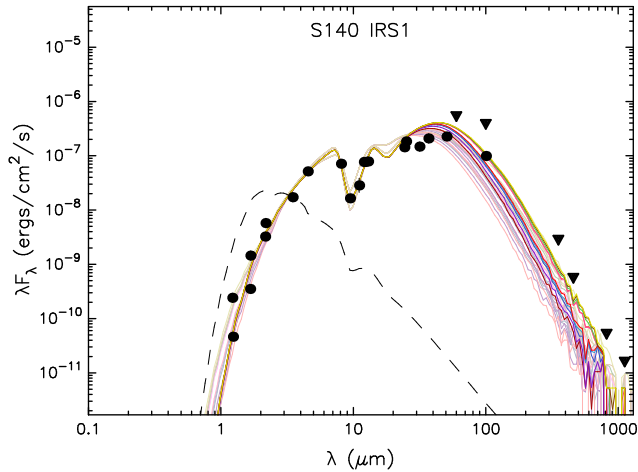}
\caption{
Fit to the SED of S140-IRS1 
made with the on-line SED fitter of Robitaille et al. (2007).
The data plotted are from 2MASS, the {\it Infrared Astronomical Satellite (IRAS)}, 
and the measurements of 
Blair et al. (1978), Lester et al. (1986), G\"urtler et al. (1991), 
Minchin et al. (1995), Mueller et al. (2002),
de Wit et al. (2009), and Harvey et al. (2012).
The dashed line shows how the best-fitting model protostar 
would look after extinction by interstellar dust but not the dust of the envelope and disc.
}
\end{figure}

\begin{figure}
\includegraphics[width=84mm]{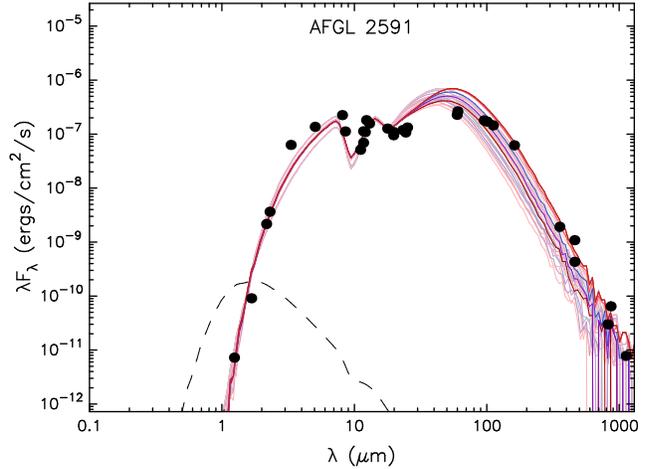}
\caption{
Fit to the SED of AFGL 2591  
made with the on-line SED fitter of Robitaille et al. (2007).
The data plotted are from 2MASS, {\it IRAS}, 
and the measurements of 
Lada et al. (1984), Jenness, Scott, \& Padman (1995), 
van der Tak et al. (1999), Marengo et al. (2000), Mueller et al. (2002), 
and de Wit et al. (2009). 
The dashed line is described in the previous figure.
}
\end{figure}

For both S140-IRS1 and AFGL 2591 the procedure for finding a good-fitting model 
is as follows: 
(1) We computed models for the top two YSO models that gave good fits to the SEDs 
(Table 5 and Figs. 9 and 10). 
As it happened, the top two models for both S140-IRS1 and AFGL 2591 have  
almost the complete range of envelope mass for each object 
and thus the differences between the models show the uncertainties of the fitted parameters.  
An interesting note is that both models of S140-IRS1 have much lower effective temperatures 
for the central star than would be expected for main sequence stars of 19 M$_\odot$. 
This could be an indication of the very young age of the S140-IRS1 protostar. 
On the other hand, both models of AFGL 2591 have high-enough effective temperatures 
that they could be ionizing their own \mbox{H\,{\sc ii}} regions; 
this is not in disagreement with the presence of ionized gas near AFGL 2591, 
which Johnston et al. (2013) suggest is a photoionized wind.

(2) We computed a series of scattering and polarization models with the inclination angle, 
THETE, varying from 10 to 40$^{\circ}$ 
(the initial inclination angle estimated by the Robitaille et al. 2007 SED fitter was 
its minimum angle, 18.19$^{\circ}$, for all four models). 
We estimate that the final inclination angle is uncertain by about 5$^{\circ}$.
For each inclination angle, models were computed with various 
cavity opening angles, THET1, and then convolved with a PSF representing 
the NICMOS 0.2 arcsec resolution. 
The best fit for THET1 was the value such that 
the ratio of the central pixels to the pixels 1 and 2 arcsec distant agreed with the data. 
Since both sources are very clumpy, the observed ratio is poorly determined,  
but it is surely large considering the observed magnitudes of 
the S140-IRS1 and AFGL 2591 protostars (Tables 3 and 4). 
For the sake of this step in the modelling procedure, we estimated 
values of this ratio of $\sim 50$ for S140-IRS1 and $\sim 100$ for AFGL 2591
from the observed surface brightness in Figs. 2(b) and 3(b). 
The resulting optical depths through the envelope to the central point are 
$\tau_{2 \mu \rm m} \sim 8.0$ for the models of S140-IRS1 
and $\sim 5.9$ for the models of AFGL 2591.

(3) Then, for each inclination angle we compared the shape of the resulting
model to the observed source and selected the final inclination angle
by the best shape. 
Because S140-IRS1, with its extensions to the north and southwest (Fig. 2b), 
almost certainly lacks the symmetry about the polar axes of the models, 
its two models were not extensively iterated for appearance. 
For Model 1 we kept the cavity opening angle of model 3000319 and only  
performed the Step 2 iteration of the inclination angle, 
and for Model 2 the inclination clearly needed to be larger than for Model 1 
with an additional adjustment of the cavity opening angle. 
The appearances in scattered light and polarization of both models are very similar. 

AFGL 2591 does not appear to have much scattered light from the cavity
(Fig. 3b) -- this was modelled by reducing the models' cavity density 
by two orders of magnitude from that of the original Robitaille et al. (2006) models.
Because of this lack of central emission, 
the shape comparison was made by minimizing the sum of the squares of the differences
of the edges of the models and source, excluding the relatively empty cavity region.
The models with the larger inclination angles have much wider opening
angles on the sky than is observed and so they gave much poorer chi-squares.

\subsection{Models of S140-IRS1}

\begin{figure}
\includegraphics[width=84mm]{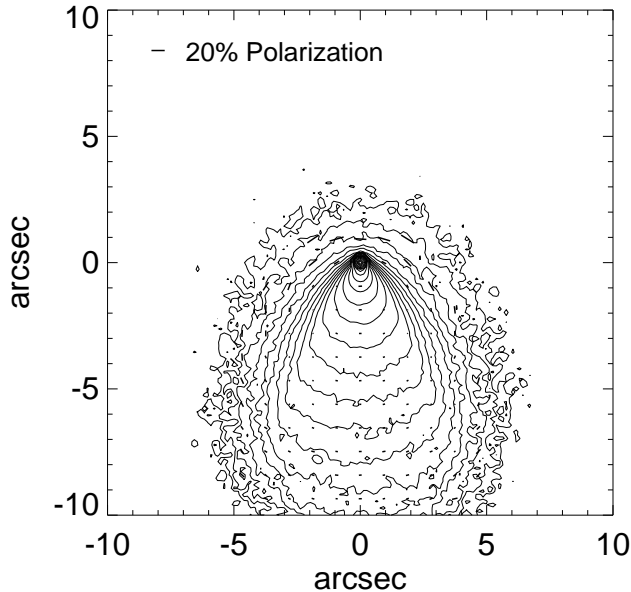}
\caption{
Monte Carlo scattering model for S140-IRS1, employing spherical grains.
The parameters used in the model are given in Table~5. 
The model is plotted with the cavity opening in the down direction 
to resemble the close-to-south-facing outflow of Fig. 6.
}
\end{figure}

Fig. 11 shows the {\sc ttsscat} Model 1 (model 3000319) for S140-IRS1 
computed with the modified parameters of Table 5.
The model is plotted with the axis of the outflow that is approaching the viewer 
at the bottom of the plot to approximate the S140-IRS1 outflow direction, 
which appears to have PA $\sim 150 - 160$\degr\ (see Fig. 6). 
We note that model 3000319 has no disc; adding a 0.5 M$_\odot$ disc to the {\sc ttsscat} model 
makes essentially no difference to the output contours and polarization vectors.

The 12\degr\ model inclination for model 3000319 (Table 5) 
is in reasonable agreement with that estimated from other observations. 
The radio observations of the CO outflow show both positive and negative velocity 
emission both north and south of IRS1 (e.g., Hayashi \& Murata 1992; Minchin et al. 1993).
Consequently, it is generally inferred that the outflow is close to the line of sight with the 
blue-shifted outflow to the southeast of IRS1 (Minchin et al. 1995; Schertl et al. 2000). 
Another indication that our line-of-sight is close to the outflow axis 
is the very bright appearance of the star at 2 \micron,
from which we infer that the line of sight does not pass through any optically thick disc. 

Our image in Fig. 2b is very wide in the direction perpendicular to the outflow,  
where we assume that the outflow direction has PA = 154\degr\ (e.g., Schertl et al. 2000).
The model in Fig. 11, however, is elongated in the direction of the outflow, 
a result of the small cavity opening angle of 5.0\degr.  
A model that has wide contours in the direction perpendicular to the outflow 
could be produced with the use of a much larger cavity opening angle; 
however, then the optical depth to the YSO would be much smaller, making the YSO too bright.
On the other hand, the narrow cavity opening angle may actually be appropriate 
because in both our data (Fig. 6) and the image of Schertl et al. (2000), 
there appears to be a narrow outflow 
in the first $\sim 3$ arcsec from the YSO ($\sim 2500$ au in the plane of the sky). 

A different approach to modelling S140-IRS1 is that of Maud et al. (2013), 
who first found a good fitting model to the NIR scattered light 
(as measured by Schertl et al. 2000)
and then used these parameters to investigate the thermal SED. 
Their resulting models are fairly similar to ours. 
However, they also point out the need for including sub-mm observations 
in the model SED fitting routines. 
We do not feel it necessary to go into such detail -- 
the models are not unique and NIR images are clearly influenced by other sources 
such as IRS3 and SMM1 (see also the images of Weigelt et al. 2002).

These models of S140-IRS1 are for only the core of the outflow, 
within about 2 -- 3 arcsec of IRS1 (e.g., Fig. 6). 
Even here the maximum polarization of the model ($\sim 10$ percent) 
is much smaller than the polarization observed in the data (Fig. 2a). 
This behaviour has been seen in other comparisons of models with polarization data 
(Simpson et al. 2009). 
For S140-IRS1, the very high polarization (maximum of $\sim 80$ percent) seen at 
great distances from IRS1 must indicate that shape of the cavity expands greatly 
to the south so that the light is scattered at close to right angles by the dust 
at $>5$ arcsec from IRS1. 
We say this because scattered light from spheres is not polarized when scattered 
in the forward direction.

However, light scattered by elongated grains can be substantially polarized even when
scattered in the forward direction (Matsumura \& Seki 1996; Whitney \& Wolff 2002).
Moreover, IRS1 itself is polarized due to 
absorption by aligned grains (absorptive polarization). 
We suggest that the aligned elongated grains are in the vicinity of IRS1 
because its polarization angle is so different from that of the other stars 
in the field of view.
We discuss aligned grains in Section 4.3.

\subsection{Models of AFGL 2591}

\begin{figure}
\includegraphics[width=84mm]{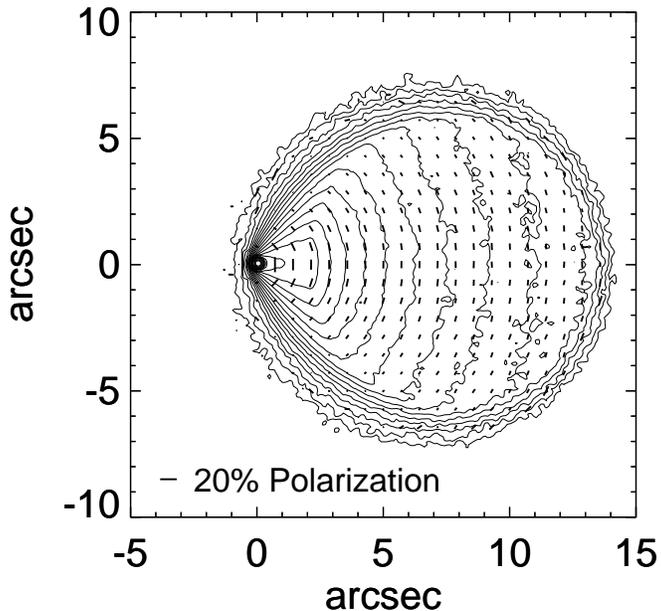}
\caption{
Monte Carlo scattering model for AFGL 2591, employing spherical grains.
The parameters used in the model are given in Table~5. 
The model is plotted with the cavity opening to the right  
to resemble the close-to-west-facing outflow of Fig. 7.
}
\end{figure}

As was done for S140-IRS1, we start with models that assume the dust grains are spherical,
which means the scattering properties of the grains can be computed with Mie theory 
and the geometry does not need to take grain alignment into account 
(Whitney \& Hartmann 1992, 1993; Stark et al. 2006).
The parameters of the best fitting SED model 3007097 (Fig. 10) are given in Table 5. 
The scattering model is the same as the SED model 
except for revised inclination and cavity opening angles -- 
this model is plotted in Fig. 12. 
The SED model inclination of 18\degr\ does not give a good fit to the observed NICMOS image --  
the {\sc ttsscat} model for model 3007097 with no adjustments to inclination or cavity opening angles
shows a monopolar scattering lobe but no illuminating star. 
This is because the optical depth through the YSO envelope is extremely large and 
the cavity opening angle is much smaller than the model inclination.
Another indicator of a poor fit is that the total extent of the outflow as projected on the sky 
is much smaller than the observed extent of the outflow.

Keeping the accretion rate and stellar parameters constant 
(since these produce the luminosity and far-infrared SED), 
we find that changing the inclination angle and cavity opening angle 
produces models with 2~\micron\ morphologies more similar 
to the appearance and polarization of AFGL 2591.
The additional change from model 3007097 is that the cavity density is reduced 
by two orders of magnitude so that there is less scattered light in the outflow near the YSO. 
With these modifications, an obvious illuminating star is present, 
with surface brightness after smoothing to the 0.2 arcsec NICMOS resolution 
that is $\sim 2$ orders of magnitude brighter than the nearby scattered flux,
in agreement with the data. 
The brightest scattered flux is located within 1 -- 2 arcsec from the illuminating star 
(we assume that the loops of scattered light at $\sim 10$ arcsec 
are due to inhomogeneities in the outflow and cavity). 

Typically these models for very young, massive YSOs have a dense envelope produced by a high 
accretion rate and an outflow cavity containing dust; some but not all models have dense discs. 
A few of the best fitting SED models for S140-IRS1 and AFGL 2591 have discs, 
such as Model 2, 3018960, for AFGL 2591. 
Tests were made by adding discs to the preferred parameters (Table 5); 
the result is that there is no appreciable difference in the scattered light output, 
probably because the toroid optical depth is so high. 
We would not distinctly detect the small, sub-Keplerian rotating disc-like structure  
that has been seen in AFGL 2591 molecular lines by Wang, van der Tak, \& Hogerheijde (2012).

The chief way in which the scattered light models do not reproduce the NICMOS 
polarization observations of AFGL 2591 seen in Fig. 7  
is that the region within 5 arcsec north and south of the YSO 
is also polarized perpendicular to the outflow (parallel to the polarization of the YSO).
This is {\it not} the same as a parallel polarization pattern, 
also known as a `polarization disc'.
Polarization discs occur when photons scatter multiple times, 
from the polar region back down to the equatorial region and 
then from the outer parts of an optically thick toroid or disc 
in the equatorial plane towards the viewer 
(e.g., Whitney \& Hartmann 1993; Simpson et al. 2009; Murakawa 2010).
Since photons are polarized perpendicular to the scattering plane, 
such photons are polarized parallel to the equatorial plane, 
giving the appearance of parallel polarization pattern.  
This is the definition of a polarization disc. 
One can see this secondarily scattered light only when 
the light scattered directly from the illuminating star
(which has a centrosymmetric polarization pattern around the illuminating star)
is so extincted as to be unimportant.
Here, the AFGL 2591 YSO is so bright that its directly scattered light dominates 
that of any light whose previous last point of scattering is in the outflow. 
Consequently, the polarized light seen north and south of the YSO 
must be that of the directly scattered light from the YSO, perhaps affected by extinction.
A polarized PSF from the YSO could also contribute polarized light to the north and south 
of the AFGL 2591 YSO in some observing modes; 
however, it does not in this case because we viewed the source 
with the YSO in the coronagraph hole and further subtracted the coronagraph PSF. 
Additional discussion of parallel polarization patterns can be found in Whitney (1995)
and Whitney, Kenyon, \& G\'omez (1997).

We will show in the next section that such polarization can occur 
from dichroic absorption by aligned grains.

\subsection{Models with non-spherical grains}

We observe substantial polarization in a large number of the stars and YSOs,
as listed in Tables 2 -- 4,
and always in the brightest objects that are YSOs.
It has long been known that stellar polarization can be caused by dichroic absorption 
by aligned grains (e.g., Martin 1974 and references therein).
In the rest of this paper we will assume that the grains are aligned by the local 
magnetic field, since such alignment is inevitable, regardless of the actual physical 
alignment mechanism (Lazarian 2007, 2009).
Here we demonstrate that the observed polarization patterns can indeed be caused by 
both dichroic absorption and scattering by magnetically aligned grains 
by computing models that have various magnetic field morphologies.

We use a radiation transfer code for scattering and extinction of aligned grains 
as described by Whitney \& Wolff (2002), and modified for arbitrary field direction 
(same website as footnote 2).
To compute the coefficients of the scattering and absorption matrices, 
we use the codes of Mishchenko, Hovenier, \& Travis (2000). 
For this work, we use dust grains that are 
prolate spheroids with a 2:1 axis ratio and a wobble of $\pm 30$\degr\ 
and compute the scattering and absorption coefficients for a wavelength of 2.2 \micron. 

Because the 3-D magnetic field geometries in collapsing protostars are not well known and still
difficult to simulate in magnetohydrodynamical (MHD) models, 
we qualitatively describe them using toy analytic formulae.
Starting with the analytic formula from Galli \& Shu (1993) from a 1-D collapse model in which 
an initially polar magnetic field is pinched by the infalling gas along the equator, 
we then add in a toroidal component assuming the field is twisted by rotation.
Thus, we write the three-dimensional $r$, $\theta$, and $\phi$ components of the magnetic field $B$ as 

\begin{equation}
B_r = {{(1+x^2)^2}\over{4x^2}} {\cos} \theta,
\end{equation}

\begin{equation}
B_\theta = -\left[{{(1+x^2)^3}\over{8x}}\right]^{1/2} {\sin} \theta,
\end{equation}

where $x$ is the radial distance from the YSO divided by two times the maximum envelope dimension, 
RMAX (see Table 5), and 

\begin{equation}
B_\phi = C_\phi {{e^{-{\rm abs}(Z/H_Z)} (B_r^2 + B_\theta^2)^{1/2}}\over{(R_{\rm cyl}/{R_C})^{1.5}}}
\end{equation}

where $C_\phi$ is a dimensionless constant, and $H_Z$ is the vertical scale height for the 
$\phi$ component of the magnetic field in au. 
$Z$ and $R_{\rm cyl}$ are the vertical (polar) and cylindrical radius, respectively, in au 
for a cylindrical coordinate system. 

Parameters for the examples in Figs. 13 and 14 are given in Table 5.
Many models were run, and it is clear that the models are not unique, because 
the parameter that has the most effect on the appearance of the polarization vectors 
is the ratio of $B_\phi/B_Z$, where $B_Z$ is the 
component of the magnetic field in the polar direction, 
and various combinations of  $C_\phi$ and $H_Z$ 
can produce the same ratio at the same position in the model. 
In Table 5, $Z1$ is the polar coordinate of the point on the cavity wall where 
$B_\phi/B_Z = 1$. 
Example contours of this ratio are plotted in Fig. 8.

For comparison with the data, we especially note that 
it is not possible to include all important details that are seen in the observed images 
in the models. 
Certainly the models are incomplete because 
the absolute values of the model polarization critically depend on the input shape and wobble 
of the grains (e.g., Whitney \& Wolff 2002), 
and that the configuration of the polarization vectors depend on the formulae in 
equations 1 -- 3. 
Moreover, numerical simulations of collapsing clouds show that the gas becomes 
very turbulent, significantly affecting the computed magnetic field (e.g., Peters et al. 2011). 
We are only demonstrating here what effects magnetic fields can have on 
models predicting polarization; 
there is otherwise too much uncertainty in the models to tightly constrain 
the details of the magnetic fields.

\subsubsection{S140-IRS1}

\begin{figure*}
\includegraphics[width=176mm]{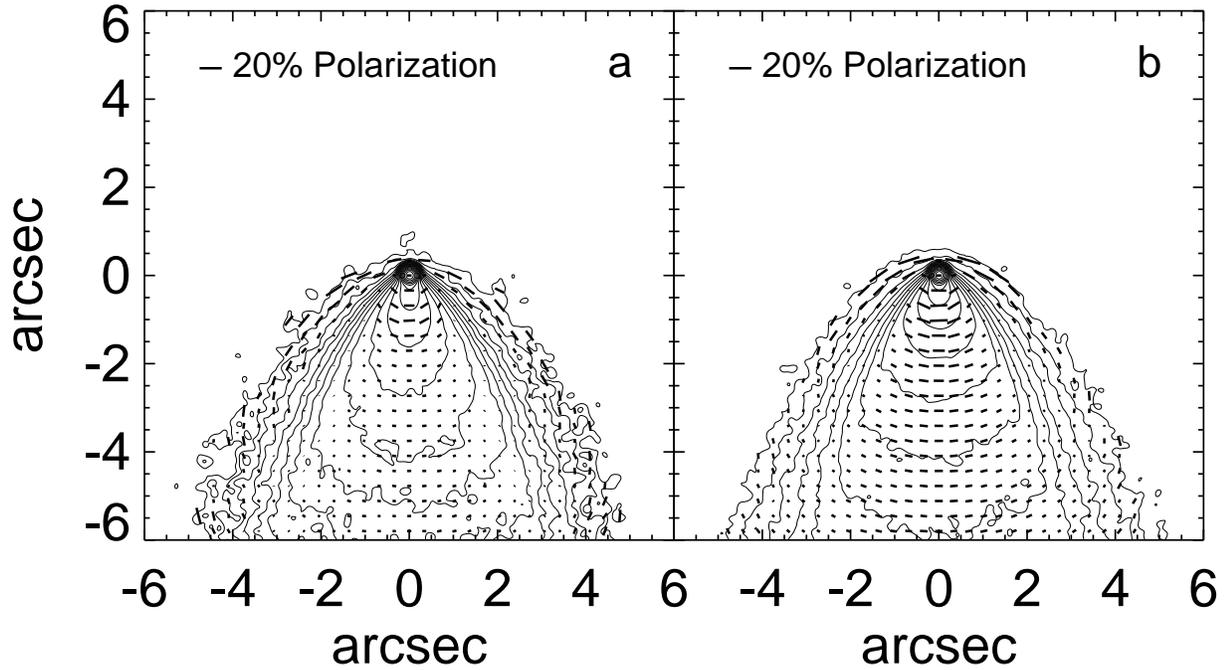}
\caption{
Monte Carlo scattering models for S140-IRS1, employing prolate, aligned grains.
The parameters used in the models are given in Table~5. 
See also Fig. 11.
Panel (a): model 1. Panel (b): model 2. 
}
\end{figure*}

Fig. 13 shows the results of the models of S140-IRS1 with parameters from Table 5. 
The overall shape of the scattered light in the outflow is very similar 
to that of the spherical-grain model in Fig. 11;
however, the aligned grain models have significantly more polarization.  
In particular, the illuminating YSO has a polarization of 4 percent, 
whereas the model with spherical grains has essentially no polarization 
at the location of the illuminating star. 
Although the amount of the YSO polarization is similar to that seen in S140-IRS1,
this is probably fortuitous because the grain axis ratio and wobble are quite arbitrary 
and could be significantly different. 
We do conclude, however, that the polarization of the illuminating YSO 
could be caused by a small toroidal magnetic field.

\subsubsection{AFGL 2591}

\begin{figure*}
\includegraphics[width=176mm]{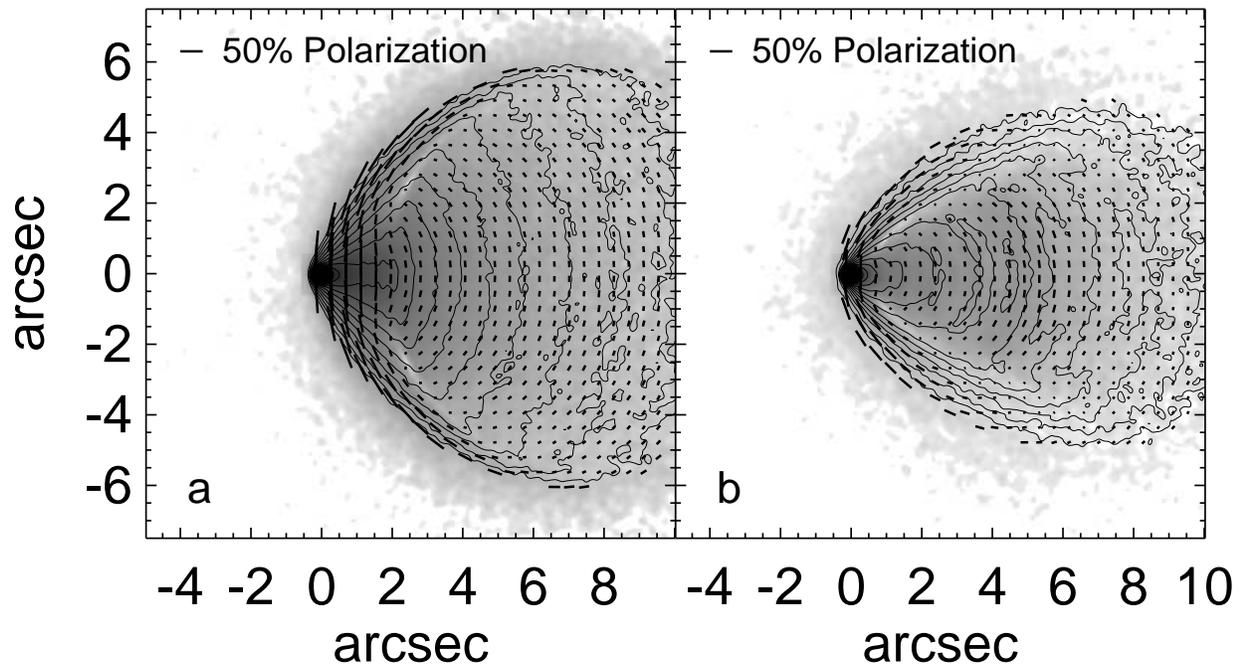}
\caption{
Monte Carlo scattering models for AFGL 2591, employing prolate, aligned grains.
The gray-scale is the polarized intensity  
and the contours are total intensity. 
The parameters used in the models are given in Table~5. 
See also Fig. 12.
Panel (a): model 1. Panel (b): model 2. 
}
\end{figure*}

Fig. 14 shows the results of the models of AFGL 2591 with parameters from Table 5.
In comparison to the spherical grain model (Fig. 12), 
this model shows substantial polarization at the YSO and all across the 
edge of the envelope that runs through the YSO. 
This aligned grain model provides a significantly improved match 
to the polarization pattern seen in the images in Fig. 7.
Neither the inclination angle nor the cavity opening angle agree exactly with 
those of models 3007097 and 3018960; 
however, the shapes overall are in reasonable agreement with that of the data 
and the angles of the polarization vectors around the outflow edges also agree 
(the inclination and cavity opening angles were estimated 
by comparing the models with the data using least squares minimization).
The two models are not identical, no doubt because of the differences in envelope mass 
and cavity opening angle.

Both models also show regions of low polarized flux on either side 
of the cavity edge, similar to the low polarized flux near the cavity edges 
that we see in Fig. 7.
Such regions are due to the switch from polarization in one direction to 
polarization different by 90\degr, resulting in very low linear polarization 
but substantial predicted circular polarization 
(up to $\pm 15$ percent for model 3007097 and up to $\pm 8$ percent for model 3018960). 
Such circular polarization was predicted by Whitney \& Wolff (2002) and Martin (1974).

We infer from this general agreement that the dust grains in AFGL 2591 are non-spherical 
and are aligned parallel to the cavity edges. 
A toroidal magnetic field is the likely cause of the alignment. 
If so, the toroidal magnetic field extends to a large fraction of the height 
of the envelope above the equatorial plane of the system. 
This is the first observational evidence that a toroidal magnetic field can extend high 
into the envelope of this young YSO. 

However, there are some disagreements between the models and the observations. 
The most obvious is that the illuminating star of the model 
has a polarization of $\sim 90$ percent 
whereas the AFGL 2591 YSO has a polarization of 16 percent (Table 3).
An easy explanation is that the actual grains are not as elongated or have more wobble 
than the model grains. 
Another possibility is that the magnetic field in AFGL 2591 is not as uniform 
as the toy magnetic field in our models. 
Both explanations are reasonable in view of the discordant appearance in the centres 
of the models and the observations, 
where the model cavity is well lit and scatters much light 
compared with the observed cavity, which is relatively dark 
but has rings of enhanced density that are probably located along the cavity outer rim
(Fig. 3).
This is in spite of the reduction of the cavity density relative to the original models. 

The models also have the configuration of their toy magnetic fields 
as toroidal within the cavity.
There are several stars, however, whose polarization vectors are more or less 
parallel to the outflow direction (Stars 2, 9, 12, and 18; Fig. 3c),
that is, in an east--west direction. 
This could be an indication that the magnetic field is affected by the outflow 
such that it becomes parallel to the cavity. 
In fact, Curran \& Chrysostomou (2007) infer that the magnetic field is parallel 
to the western part of the AFGL 2591 outflow from sub-mm polarimetry 
made with $\sim 14$ arcsec beam. 
Testing this would require a more elaborate magnetic field geometry and 
cavity density distribution than is possible with our current models.

\section{Discussion}

We have shown that all four of the massive YSOs in our study 
and the two massive YSOs in S255-IRS1 (Simpson et al. 2009) 
have polarization vectors perpendicular to their outflows, 
and for the two sources that we have modelled, 
the polarization is consistent with what would be caused by dichroic absorption 
by grains aligned by a toroidal magnetic field. 
The MIR polarization PAs, almost certainly due to dichroic absorption
considering their deep 10 \micron\ silicate absorption features, 
are $\sim 10$ and $\sim 170^\circ$ for S140-IRS1 and AFGL 2591, respectively
(Smith et al. 2000).
Although the agreement of the NIR and MIR polarizations is excellent for AFGL 2591, 
the MIR polarization PA of S140-IRS1 differs from the NIR value by $\sim 30^\circ$. 
It may be that the envelope dust model shown in Fig. 8 is not completely valid --  
the $\sim 10^\circ$ polarization PA at 10 \micron\ could indicate 
that the magnetic field direction in the outer envelope of S140-IRS1 
is not uniformly toroidal or the dust is not uniform, 
causing the apparent rotation of the PA. 

We note that all the sources that we have studied are high-mass YSOs. 
Hull et al. (2013) have performed high-spatial-resolution (2 -- 3 arcsec) mm wavelength interferometry 
of a number of low-mass YSOs within 415 pc of the Earth. 
Their goal was to determine if the mm polarization is aligned with the YSO outflows; 
they conclude that it is not and is consistent with being random.
They also point out that if the magnetic field becomes wrapped around the outflow axis 
by rotation, such an intrinsic toroidal field could appear random 
as a result of averaging along the line of sight and averaging over 
their 2.5 arcsec beams. 
We notice that several of their sources with plotted polarization vectors 
have the polarization perpendicular to the outflow direction. 

There are low-mass sources that reveal evidence of aligned grains through NIR polarimetry. 
For example, Lucas \& Roche (1998) observed and modelled a number of low-mass YSOs;
several of their sources require aligned grains for the models to agree with their observations. 
Rodgers et al. (E. Rodgers, in preparation) are also modelling the low-mass YSOs 
whose polarization was observed with {\it HST} NICMOS.
Their models include either spherical grains with a wide variety of grain compositions 
or aligned elongated grains. 
As is the case with our data, they find that the observed YSOs have much higher polarization 
than can be produced with any models using spherical grains. 
However, they also find that models using aligned grains 
and also models where the dust is highly clumped 
give much better agreement with their observations.

Rotation of the polarization PA within a source produces 
not only regions of low polarization percentage, but also strong circular polarization 
(e.g., Martin 1974; Whitney \& Wolff 2002). 
Lonsdale et al. (1980) measured statistically significant circular polarization in both 
AFGL 2591 ($-0.85 \pm 0.08$ percent) and S140-IRS1 ($-0.93 \pm 0.12$ percent) in a 10-arcsec beam.
As expected from the models of Whitney \& Wolff (2002), 
our models show negligible circular polarization at the position of the star 
(nor integrated circular polarization because the positive and negative components are 
almost equal).
However, the models show substantial circular polarization 1 -- 3 arcsec offset from the star.
As described previously, this is due to the rotation of the PA of the 
polarization vectors as one goes from the centrosymmetric polarization vectors of 
scattering to the absorptive polarization indicating the PA of the 
magnetic field (e.g., Martin 1974; Whitney \& Wolff 2002).
Clearly high spatial resolution mapping of the circular polarization would be desirable 
to test for these predicted effects of aligned grains. 
Moreover, the occurrence of statistically significant circular polarization in the 10-arcsec beam 
of Lonsdale et al. (1980) indicates that there are spatial asymmetries in the 
observed sources that are not included in the models. 
An example of an asymmetry that could be ascertained from high-resolution circular polarimetry is
the helical magnetic field that has been inferred from the circular polarization observations 
of HH 135-136 by Chrysostomou, Lucas, \& Hough (2007).

The recent incorporation of magnetic fields into numerical simulations 
of the formation of massive stars 
promises to significantly advance our understanding of these objects. 
Although it is generally thought that magnetic fields cannot prevent the formation 
of a massive star (e.g., McKee \& Ostriker 2007),
the presence of a magnetic field changes the details of the collapse
(Klessen, Krumholz, \& Heitsch 2011). 
One question is whether a disc can form if magnetic fields brake the regular flow of gas
through the disc to the star -- 
the results from various simulations differ in their conclusions 
depending on specific assumptions about both the microphysics 
and the strength and initial orientation of the magnetic fields 
(e.g., Mellon \& Li 2008, 2009; Joos, Hennebelle, \& Ciardi 2012; Seifried et al. 2012; 
Li, Krasnopolsky, \& Shang 2013). 
We can not test this aspect of massive star formation because 
our models are not particularly sensitive to the disc parameters since
the massive envelopes dominate the images and SEDs.

In all the recent MHD simulations, 
rotation causes the magnetic field lines to become wound around the central condensation,
in effect, to become a toroidal magnetic field. 
Examples include the simulations of Hennebelle \& Fromang (2008), 
Peters et al. (2011) and Seifried et al. (2011). 
These authors simulated the formation of a star 
from a rotating cloud that initially also contained a polar magnetic field.
With time the gas in the centre becomes very turbulent 
causing the magnetic field lines to appear disorganized, 
but with a strong toroidal component. 
The toroidal component extends high along the rotation axis of the simulated model,
similar to the `magnetic tower' described by Lynden-Bell (1996).
Our toy magnetic geometries have similarities to the magnetic structures in these simulations;
however, our geometries are much more uniform and regular. 
Almost certainly the grains in real YSO envelopes, if the gas and dust are as turbulent 
as these simulations, would not be as aligned as the grains in our models.
This is a likely reason why our models of AFGL 2591 have so much more absorptive polarization 
than is observed.

\section{Summary and conclusions}

We have measured the 2 \micron\ polarization in three 
well-studied sources, Mon R2-IRS3, S140-IRS1, and AFGL 2591
with NICMOS on {\it HST}. 
Mon R2-IRS3 contains at least two YSOs with monopolar outflows 
and S140-IRS1 and AFGL 2591 each have a single outflow; 
all of these outflows show substantial polarization.
The illuminating stars of the outflows are also polarized 
in a direction perpendicular to the outflows; 
we attribute this polarization to absorption by aligned grains.

Numerous stars in the field of view of each YSO were also observed.
The polarization was measured for the stars, and in most cases, the polarization 
vectors do not align with the polarization of the YSOs or their outflows.
We suspect that most of the stars with very low polarization are foreground 
to the YSOs.
On the other hand, most of the stars with significant polarization 
are probably members of the cluster of stars forming around each of the massive YSOs. 

We have modelled the scattered light for S140-IRS1 and AFGL 2591 
using both spherical grains and elongated grains that are aligned with 
proposed magnetic field orientations. 
Only the models with aligned grains can produce polarization 
at the position of the star that is illuminating the diffuse nebulosity. 
Our toy magnetic field geometry can be described as 
a polar field with an equatorial pinch with the addition of a possible 
or even substantial toroidal component. 
The models that produce polarization 
at the correct PAs (perpendicular to the outflow cavities) 
all have a substantial toroidal component to the magnetic field 
in the equatorial plane, perpendicular to the outflow. 
The toroidal magnetic field in the model that best fits AFGL 2591 
extends to a substantial fraction of the height of the model cavity, which is $10^5$ au.
We conclude that the morphologies of all the massive YSOs in this study 
are consistent with the presence of a toroidal magnetic field 
and the toroidal component of the field in the most massive of the objects, AFGL 2591, 
extends high into the envelope. 

\section*{Acknowledgements}
We thank the referee for his careful comments, which greatly improved the presentation.
Support for programme 10519 was provided by NASA through a grant from
the Space Telescope Science Institute, which is operated by the Association of Universities
for Research in Astronomy, under NASA contract NAS5-26555.

\bsp

\label{lastpage}

\end{document}